\documentclass[review]{elsarticle}
\usepackage{pgfplots}
\pgfplotsset{compat=newest}
\pgfplotsset{plot coordinates/math parser=false}
\newlength\fheight
\newlength\fwidth
\usepackage{color,soul}
\usepackage{hyperref}
\usepackage{amsmath}
\usepackage{amsfonts}
\usepackage{amssymb}
\usepackage{flexisym}
\usepackage{graphicx}
\usepackage{epstopdf} 
\usepackage[margin=2.5cm]{geometry}                           
\usepackage[font=normalsize,labelfont=bf,labelsep=space,skip=2pt,figurename=Fig.]{caption}	
\usepackage{color} 											  
\usepackage{float}											  
\usepackage{subcaption}													
\usepackage{multirow}													
\usepackage{array}														
\usepackage{booktabs}													
\usepackage{lmodern}													
\usepackage{algorithm}
\usepackage{algpseudocode} 												
\newdefinition{rmk}{Remark}

\usepackage{tikz}														%
\usetikzlibrary{matrix,shapes,arrows,positioning,chains,calc,intersections}				%
\usepackage{placeins}
\newcommand{\ra}[1]{\renewcommand{\arraystretch}{#1}}							
\biboptions{sort&compress}				
\makeatletter  															
\def\ps@pprintTitle{%
	\let\@oddhead\@empty
	\let\@evenhead\@empty
	\def\@oddfoot{}%
	\let\@evenfoot\@oddfoot}
\makeatother

\journal{Journal of \LaTeX\ Templates}

\begin{document}

\begin{frontmatter}

\title{Using Approximate Bayesian Computation by Subset Simulation for Efficient Posterior Assessment of Dynamic State-Space Model Classes}

\author[mymainaddress,mysecondaryaddress]{Majid K. Vakilzadeh}

\author[mymainaddress]{James L. Beck\corref{mycorrespondingauthor}}
\cortext[mycorrespondingauthor]{Corresponding author}
\ead{jimbeck@caltech.edu}

\author[mysecondaryaddress]{Thomas Abrahamsson}

\address[mymainaddress]{Division of Engineering and Applied Science, California Institute of Technology, CA, USA}
\address[mysecondaryaddress]{Department of Applied Mechanics, Chalmers University of Technology, Gothenburg, Sweden}

\begin{abstract}
Approximate Bayesian Computation (ABC) methods have gained in their popularity over the last decade because they expand the horizon of Bayesian parameter inference methods to the range of models for which only forward simulation is available. The majority of the ABC methods rely on the choice of a set of summary statistics to reduce the dimension of the data. However, as has been noted in the ABC literature, the lack of convergence guarantees that is induced by the absence of a vector of sufficient summary statistics that assures inter-model sufficiency over the set of competing models, hinders the use of the usual ABC methods when applied to Bayesian model selection or assessment. In this paper, we present a novel ABC model selection procedure for dynamical systems based on a newly appeared multi-level Markov chain Monte Carlo method, self-regulating ABC-SubSim, and a hierarchical state-space formulation of dynamic models. We show that this formulation makes it possible to independently approximate the model evidence required for assessing the posterior probability of each of the competing models. We also show that ABC-SubSim not only provides an estimate of the model evidence as a simple by-product but also it gives the posterior probability of each model as a function of the tolerance level, which allows the ABC model choices made in previous studies to be understood. We illustrate the performance of the proposed framework for ABC model updating and model class selection by applying it to two problems in Bayesian system identification: a single degree-of-freedom bilinear hysteretic oscillator and a three-story shear building with Masing hysteresis, both of which are subject to a seismic excitation. 
       
\end{abstract}


\end{frontmatter}

\section{Introduction}
In many areas of science such as biology, economics, social sciences and engineering, it is desired to make inference about the parameters of a mathematical model based on the experimental data from a real system in order to make more accurate predictions of the system behavior and to better understand it. Furthermore, there are invariably multiple candidate models with different mathematical forms to represent the system behavior and so there is a need to assess their plausibility based on the experiment data. The fully probabilistic Bayesian approach provides a rigorous framework to achieve these goals while also properly quantifying the uncertainty in the model parameters induced by uncertainty in the measurements and the accuracy of the mathematical model. In the Bayesian approach, a key idea is to construct a \textit{stochastic model class} $\mathcal{M}$ consisting of the following fundamental probability distributions \cite{beck2010bayesian}: a set of parameterized input-output probability models $p(\boldsymbol{y}|\boldsymbol{\theta},\boldsymbol{u},\mathcal{M})$ for predicting the system behavior of interest $\boldsymbol{y}$ for given input $\boldsymbol{u}$ and a \textit{prior} probability density function (PDF) $p(\boldsymbol{\theta}|\mathcal{M})$ over the parameter space $\boldsymbol{\Theta}\in\mathbb{R}^{N_p}$ of $\mathcal{M}$ that reflects the relative degree of plausibility of each input-output model in the set. When data $\mathcal{D}$ consisting of the measured system input $\hat{\boldsymbol{u}}$ and output $\hat{\boldsymbol{z}}$ are available, the prior PDF $p(\boldsymbol{\theta}|\mathcal{M})$ can be updated through Bayes' Theorem to obtain the \textit{posterior} PDF for the uncertain model parameters $\boldsymbol{\theta}$ as: 
\begin{equation}\label{eq:eq1}
	p(\boldsymbol{\theta}|\mathcal{D},\mathcal{M})=\frac{p(\hat{\boldsymbol{z}}|\boldsymbol{\theta},\hat{\boldsymbol{u}},\mathcal{M})p(\boldsymbol{\theta}|\mathcal{M})}{p(\hat{\boldsymbol{z}}|\hat{\boldsymbol{u}},\mathcal{M})}\propto p(\hat{\boldsymbol{z}}|\boldsymbol{\theta},\hat{\boldsymbol{u}},\mathcal{M})p(\boldsymbol{\theta}|\mathcal{M})
\end{equation}
where $p(\hat{\boldsymbol{z}}|\boldsymbol{\theta},\hat{\boldsymbol{u}},\mathcal{M})$ denotes the \textit{likelihood function} of $\boldsymbol{\theta}$ which gives the probability of getting the data based on the input-output probability model $p(\boldsymbol{y}|\boldsymbol{\theta},\boldsymbol{u},\mathcal{M})$ and $p(\hat{\boldsymbol{z}}|\hat{\boldsymbol{u}},\mathcal{M})=\int_{\boldsymbol{\Theta}}\,p(\hat{\boldsymbol{z}}|\boldsymbol{\theta},\hat{\boldsymbol{u}},\mathcal{M})\allowbreak\,p(\boldsymbol{\theta}|\mathcal{M})\,d\boldsymbol{\theta}$ denotes the \textit{evidence}, or marginal likelihood, for model class $\mathcal{M}$. Despite the fact that $p(\hat{\boldsymbol{z}}|\hat{\boldsymbol{u}},\mathcal{M})$ is a constant and does not affect the shape of the posterior PDF, it is well known that it plays a crucial role in model class selection and averaging (e.g., \cite{beck2010bayesian,beck2004model,cheung2010calculation}). 

If $\boldsymbol{M}\equiv\{\mathcal{M}_1,\,\mathcal{M}_2,\,\dots,\,\mathcal{M}_{L}\}$ is a set of competing candidate model classes for a real system, then the posterior probability of each model class is given by Bayes' Theorem at the model-class level:
\begin{equation}\label{eq:eq1_1}
p(\mathcal{M}_l|\mathcal{D},\boldsymbol{M})\propto p(\hat{\boldsymbol{z}}|\hat{\boldsymbol{u}},\mathcal{M}_l)\,p(\mathcal{M}_l|\boldsymbol{M}),\,\,\,\, l=1,\,\dots,\,L
\end{equation}
in which $p(\mathcal{M}_l|\boldsymbol{M})$ denotes the prior probability of the model class $\mathcal{M}_l$. This posterior distribution quantifies the plausibility of each $\mathcal{M}_l$ to represent the uncertain behavior of the real system. If there are multiple model parameters treated as continuous stochastic variables for $\mathcal{M}_l$ then calculation of its evidence, $p(\hat{\boldsymbol{z}}|\hat{\boldsymbol{u}},\mathcal{M}_l)$, involves evaluation of a high dimensional integral over the parameter space that is computationally prohibitive. In addition, there are some model classes, e.g., hidden Markov models or dynamical state-space models, for which the likelihood function is difficult or even impossible to compute, but one might still be interested to perform Bayesian parameter inference or model selection. Approximate Bayesian Computation (ABC) methods were originally conceived to circumvent the need for computation of the likelihood by simulating samples from the corresponding input-output probability model $p(\boldsymbol{y}|\boldsymbol{\theta},\boldsymbol{u},\mathcal{M})$. 

The basic idea behind ABC is to avoid evaluation of the likelihood function in the posterior PDF $p(\boldsymbol{\theta}|\mathcal{D},\mathcal{M})\propto\,p(\hat{\boldsymbol{z}}|\boldsymbol{\theta},\hat{\boldsymbol{u}},\mathcal{M})\,p(\boldsymbol{\theta}|\mathcal{M})$ over the parameter space $\boldsymbol{\theta}$ by using an augmented posterior PDF:
\begin{equation}\label{eq:eq2}
	p(\boldsymbol{\theta},\boldsymbol{y}|\mathcal{D},\mathcal{M})\propto P(\hat{\boldsymbol{z}}|\boldsymbol{y},\boldsymbol{\theta})\,p(\boldsymbol{y}|\boldsymbol{\theta},\hat{\boldsymbol{u}},\mathcal{M})\,p(\boldsymbol{\theta}|\mathcal{M})
\end{equation}
over the joint space of the model parameters $\boldsymbol{\theta}$ and the model output $\boldsymbol{y}$ that is simulated using the distribution $p(\boldsymbol{y}|\boldsymbol{\theta},\hat{\boldsymbol{u}},\mathcal{M})$. The interesting point of this formulation is the degree of freedom brought by the choice of function $P(\hat{\boldsymbol{z}}|\boldsymbol{y},\boldsymbol{\theta})$. The original ABC algorithm \cite{tavare1997inferring} defines $P(\hat{\boldsymbol{z}}|\boldsymbol{y},\boldsymbol{\theta})=\delta_{\hat{\boldsymbol{z}}}(\boldsymbol{y})$, where $\delta_{\hat{\boldsymbol{z}}}(\boldsymbol{y})$ is equal to $1$ when $\hat{\boldsymbol{z}}=\boldsymbol{y}$ and equal to $0$ otherwise, to retrieve the target posterior distribution when $\boldsymbol{y}$ exactly matches $\hat{\boldsymbol{z}}$. However, the probability of generating exactly $\hat{\boldsymbol{z}}=\boldsymbol{y}$ is zero for continuous stochastic variables. 

Pitchard et al. \cite{pritchard1999population} broadened the realm of the applications for which ABC algorithm can be used by replacing the point mass at the observed output data $\hat{\boldsymbol{z}}$ with an indicator function $\mathbb{I}_{S(\epsilon)}(\boldsymbol{y})$, where $\mathbb{I}_{S(\epsilon)}(\boldsymbol{y})$ gives $1$ over the set $S(\epsilon)=\{\boldsymbol{y}:\rho(\boldsymbol{\eta}(\hat{\boldsymbol{z}})-\boldsymbol{\eta}(\boldsymbol{y}))\leq\epsilon\}$ and $0$ elsewhere, for some chosen metric $\rho$ and low-dimensional summary statistic $\boldsymbol{\eta}$. In this case, the approximate posterior PDF can be written as:
\begin{equation}\label{eq:eq3}
	p(\boldsymbol{\theta},\boldsymbol{y}|\mathcal{D},\epsilon,\mathcal{M})\propto \mathbb{I}_{S(\epsilon)}(\boldsymbol{y})\,p(\boldsymbol{y}|\boldsymbol{\theta},\hat{\boldsymbol{u}},\mathcal{M})\,p(\boldsymbol{\theta}|\mathcal{M})
\end{equation}
In this manner, Pitchard et al. \cite{pritchard1999population} imposed two layers of approximation on the target posterior PDF $p(\boldsymbol{\theta},\boldsymbol{y}|\mathcal{D},\epsilon,\mathcal{M})$: (a) a tolerance parameter $\epsilon$ to assign a non-zero probability only for a region in $(\boldsymbol{\theta},\boldsymbol{y})$ space where $\boldsymbol{y}$ closely approximates $\hat{\boldsymbol{z}}$, i.e., $\hat{\boldsymbol{z}}\approx\boldsymbol{y}$; and (b) summary statistics $\boldsymbol{\eta}(.)$ to summarize data in a low-dimensional space giving a weak form of the comparisons. If the summary statistics are sufficient for identification of $\boldsymbol{\theta}$, the approximation error vanishes, that is, $p(\boldsymbol{\theta},\boldsymbol{y}|\mathcal{D},\epsilon,\mathcal{M})=p(\boldsymbol{\theta},\boldsymbol{y}|\mathcal{D},\mathcal{M})$ as $\epsilon\rightarrow0$. Otherwise, summarizing the data induces a second level of approximation of the target posterior distribution. Algorithm \ref{algorithm:1} gives a pseudo-code to draw $J$ samples from the approximate posterior distribution $p(\boldsymbol{\theta},\boldsymbol{y}|\mathcal{D},\epsilon,\mathcal{M})$. 

\begin{algorithm}[htbp!]
	\caption{Standard ABC rejection algorithm \cite{tavare1997inferring}}\label{algorithm:1}
	\begin{algorithmic}
		\For {$j=1$ \textbf{to} $J$}
		\While {$\rho(\boldsymbol{\eta}(\boldsymbol{y}\textprime),\boldsymbol{\eta}(\hat{\boldsymbol{z}}))>\epsilon$} 
		\State Draw a candidate sample $\boldsymbol{\theta}\textprime\sim p(\boldsymbol{\theta}|\mathcal{M})$.
		\State Generate $\boldsymbol{y}\textprime\sim p(\boldsymbol{y}|\boldsymbol{\theta}\textprime,\mathcal{M})$. 
		\EndWhile   
		\State Set $(\boldsymbol{\theta}^{(j)},\boldsymbol{y}^{(j)})=(\boldsymbol{\theta}\textprime,\boldsymbol{y}\textprime)$.
		\EndFor	
	\end{algorithmic}
\end{algorithm}

Algorithm \ref{algorithm:1} gives samples from the true posterior distribution when the tolerance parameter $\epsilon$ is sufficiently small and the summary statistics $\boldsymbol{\eta}(.)$ are sufficient. These conditions pose some difficulties for computer implementation of this algorithm which renders it far from a routine use for parameter inference and model selection. Firstly, a sufficiently small tolerance parameter $\epsilon$ means that only predicted model outputs $\boldsymbol{y}$ lying in a small local neighborhood centered on the observed data vector $\hat{\boldsymbol{z}}$ are accepted. However, this leads to a problem of rare-event simulation and so if Algorithm \ref{algorithm:1} is used, the model output $\boldsymbol{y}$ must be computed for a huge number of candidate samples in order to produce an acceptable sample size in the data-approximating region $S(\epsilon)$. Thus, many ABC algorithms have emerged to enhance the computational efficiency of the basic ABC rejection algorithm, e.g., ABC-MCMC \cite{bortot2007inference,marjoram2003markov,sisson2011likelihood} , ABC-PRC \cite{sisson2008note,sisson2007sequential}, ABC-SMC \cite{del2012adaptive, drovandi2011estimation,toni2009approximate}, ABC-PMC \cite{beaumont2009adaptive} and ABC-SubSim \cite{chiachio2014approximate}. 

Secondly, the lack of a reasonable vector of summary statistics that works across models hinders the use of an ABC algorithm for model selection \cite{didelot2011likelihood}. The ABC solution to the Bayesian model selection problem is to perform the inference at a model class level by incorporating a model index within the model parameters. In this approach, a prior  distribution $p(\mathcal{M}_l)$ for $l=1,\,\dots,\,L$ is assigned to the competing models in the set $\boldsymbol{M}\equiv\{\mathcal{M}_1,\,\mathcal{M}_2,\,\dots,\,\mathcal{M}_{L}\}$ along with a prior distribution $p(\boldsymbol{\theta}|\mathcal{M}_l)$ for the parameters conditional on the model index $\mathcal{M}_l$. Then, following the standard ABC rejection algorithm at the model class level, as given by the pseudo-code in Algorithm \ref{algorithm:2}, the posterior probability of each of the candidate models can be readily estimated as follows:
\begin{equation}\label{eq:eq4}
	p(\mathcal{M}_l|\hat{\boldsymbol{z}},\epsilon)\approx \frac{1}{J}\sum_{j=1}^{J}\mathbb{I}_{m^{(j)}=\mathcal{M}_l},\,\,\,\,\mbox{for}\,\, l=1,\,\dots,\,L 
\end{equation}
which is basically the frequency of acceptance from model $\mathcal{M}_l$. Let the vector $\boldsymbol{\eta}_{\boldsymbol{M}}(\boldsymbol{y})=[\boldsymbol{\eta}_1^T(\boldsymbol{y}),\,\dots,\,\boldsymbol{\eta}_{L}^T(\boldsymbol{y})]^T$ denote the concatenation of the summary statistics used for all models. Grelaud et al. \cite{grelaud2009abc} reported that sufficiency of the summary statistics $\boldsymbol{\eta}_l(\boldsymbol{y})$ for models in the set $\boldsymbol{M}$ does not guarantee sufficiency of $\boldsymbol{\eta}_{\boldsymbol{M}}(\boldsymbol{y})$ for comparison of those models. In fact, forming sufficient statistics for model comparison is not feasible in most problems for which ABC model selection has been implemented \cite{robert2011lack}. In this setting, the approximations made by Algorithm 2 does not always converge to the true model posterior probability \cite{roberts2009examples}.

Note that one can resort to entire data in Algorithm \ref{algorithm:2} instead of using the summary statistics. This avoids a loss of information in the metric $\rho(\boldsymbol{\eta}_{\boldsymbol{M}}(\boldsymbol{y}\textprime),\boldsymbol{\eta}_{\boldsymbol{M}}(\hat{\boldsymbol{z}}))$ and, in turn, leads to a consistent decision for model choice \cite{robert2011lack}. Toni et al. \cite{toni2009approximate} and Toni and Stumpf \cite{toni2010simulation} developed an ABC algorithm based on Sequential Monte Carlo (ABC-SMC) and modified it such that a model index is incorporated within the model parameters. The posterior probability for each model class is then estimated based on the distance between the entire measured and simulated output. However, the estimated model posterior probabilities by ABC-SMC are affected by the choice of the tolerance parameter $\epsilon$ and the variance of the proposal distribution of the Markov chain used in ABC-SMC, which imposes a type of dependency on the estimates of $p(\mathcal{M}_l|\hat{\boldsymbol{z}},\epsilon)$ that is irrelevant to the statistical problem under investigation \cite{marin2012approximate}.

\begin{algorithm}[htbp!]
	\caption{Standard ABC algorithm for model comparison \cite{marin2012approximate}}\label{algorithm:2}
	\begin{algorithmic}
		\For {$j=1$ \textbf{to} $J$}
		\Repeat 
		\State Draw $\mathcal{M}\textprime$ from the prior distribution $p(\mathcal{M}_l)$ for $l=1,\,\dots,\,L$.
		\State Draw a candidate sample $\boldsymbol{\theta}\textprime\sim p(\boldsymbol{\theta}|\mathcal{M}\textprime)$. 
		\State Generate $\boldsymbol{y}\textprime\sim p(\boldsymbol{y}|\boldsymbol{\theta}\textprime,\mathcal{M}\textprime)$.
		\Until {$\rho(\boldsymbol{\eta}_{\boldsymbol{M}}(\boldsymbol{y}\textprime),\boldsymbol{\eta}_{\boldsymbol{M}}(\hat{\boldsymbol{z}}))\leq\epsilon$}   
		\State Set $m^{(j)}=\mathcal{M}\textprime$ and $(\boldsymbol{\theta}^{(j)},\boldsymbol{y}^{(j)})=(\boldsymbol{\theta}\textprime,\boldsymbol{y}\textprime)$.
		\EndFor	
	\end{algorithmic}
\end{algorithm}

Wilkinson \cite{wilkinson2008bayesian,wilkinson2013approximate} proposed to replace the indicator function $\mathbb{I}_{S(\epsilon)}(\boldsymbol{y})$ over the data approximating region $S(\epsilon)$ with a probability distribution function $p(\hat{\boldsymbol{z}}|\boldsymbol{y},\epsilon)$, centered at $\boldsymbol{y}$, to obtain the following approximate posterior distribution:
\begin{equation}\label{eq:eq5}
	p(\boldsymbol{\theta},\boldsymbol{y}|\mathcal{D},\epsilon,\mathcal{M})\propto p(\hat{\boldsymbol{z}}|\boldsymbol{y},\epsilon)\,p(\boldsymbol{y}|\boldsymbol{\theta},\hat{\boldsymbol{u}},\mathcal{M})\,p(\boldsymbol{\theta}|\mathcal{M})
\end{equation}
This suggests a departure from the previous perspective so that $p(\boldsymbol{\theta},\boldsymbol{y}|\hat{\boldsymbol{z}},\mathcal{M})$ is now interpreted as an exact posterior PDF for a new model in which the summary statistics are corrupted by a uniform error. Didelot et al. \cite{didelot2011likelihood} showed that if one formulates the posterior distribution in (\ref{eq:eq5}) using the entire data, its normalizing constant converges to the marginal likelihood $p(\hat{\boldsymbol{z}}|\mathcal{M})$ as $\epsilon\rightarrow0$. 

In this study, we show that formulating a dynamical system as a general hierarchical state-space model enables us to independently estimate the model evidence for each model class. The recently proposed multi-level MCMC algorithm called Approximate Bayesian Computation by Subset Simulation (ABC-SubSim), is applied to solve the Bayesian inference problem of the uncertain parameters of the stochastic state-space model. We show that not only can the model evidence be estimated as a by-product of the ABC-SubSim algorithm, but also using the MCMC samples one can estimate the probability that model output $\boldsymbol{y}$ falls into the data-approximating region $S(\epsilon)$ as a function of $\epsilon$. The inherent difficulty of the ABC method for estimation of the parameters of the uncertain prediction error for stochastic state-space model is addressed and a new solution based on Laplace's method of asymptotic approximation is presented. The effectiveness of the ABC-SubSim algorithm for Bayesian model updating and model class selection with simulated data is illustrated using two Bayesian system identification examples selected from the literature: (\textit{i}) a single degree-of-freedom bilinear hysteretic oscillator \cite{beck2004model} for which the true system is not included in the set of competing models; and (\textit{ii}) a three-story shear building with Masing hysteresis \cite{muto2008bayesian}, both of which are subject to seismic excitation. These numerical examples demonstrate the performance of ABC-SubSim for solving the Bayesian model updating and model class selection problem for dynamic models with a relatively large parameter space.

\section{Formulation}\label{sec:for}
In this section, we review the formulation of a Bayesian hierarchical model class for dynamical systems and then we employ the recently-appeared algorithm for Bayesian updating, Approximate Bayesian Computation by Subset Simulation (ABC-SubSim) \cite{chiachio2014approximate} to explore its posterior PDF. We finally address the Bayesian model selection approach for the hierarchical stochastic state-space models.

\subsection{Formulation of hierarchical stochastic model class}
We construct a hierarchical stochastic state-space model class $\mathcal{M}(\epsilon)$ to predict the uncertain input-output behavior of a system. The reason for the dependence on a parameter $\epsilon$ will become evident later in this section. 

We start with the general case of a discrete-time finite-dimensional state-space model of a real dynamic system:
\begin{equation}\label{eq:eq6}
	\begin{matrix}
		\begin{split}
			\forall n\in\mathbb{Z}^+, \,\,&\boldsymbol{x}_n=\boldsymbol{f}_n(\boldsymbol{x}_{n-1},\boldsymbol{u}_{n-1},\boldsymbol{\theta}_s)\\
			&\boldsymbol{y}_n=\boldsymbol{g}_n(\boldsymbol{x}_n,\boldsymbol{u}_n,\boldsymbol{\theta}_s)
		\end{split}
		&
		\begin{split}
			(\mbox{State evolution})\\
			(\mbox{Output})
		\end{split}
	\end{matrix}
\end{equation}
where $\boldsymbol{u}_n\in \mathbb{R}^{N_I}$, $\boldsymbol{x}_n\in \mathbb{R}^{N_s}$ and $\boldsymbol{y}_n\in \mathbb{R}^{N_o}$ denote the (external) input, model state and output vector at time $t_n$, and $\boldsymbol{\theta}_s\in \mathbb{R}^{N_p}$ is a vector of uncertain-valued model parameters. For the general case of stochastic embedding, we introduce the uncertain state and output prediction errors $\boldsymbol{w}_n$ and $\boldsymbol{v}_n$ into this deterministic model to account for the model being always an approximation of the real system behavior, regardless of the choice of $\boldsymbol{\theta}_s$ \cite{beck2010bayesian}:
\begin{equation}\label{eq:eq7}
	\begin{split}
		\forall n\in\mathbb{Z}^+, \,\,&\boldsymbol{x}_n=\boldsymbol{f}_n(\boldsymbol{x}_{n-1},\boldsymbol{u}_{n-1},\boldsymbol{\theta}_s)+\boldsymbol{w}_n\\
		&\boldsymbol{y}_n=\boldsymbol{g}_n(\boldsymbol{x}_n,\boldsymbol{u}_n,\boldsymbol{\theta}_s)+\boldsymbol{v}_n
	\end{split}
\end{equation}
where we now re-define $\boldsymbol{x}_n$ and $\boldsymbol{y}_n$ to be the dynamic system state and output vectors at time $t_n$, not the model state and output. The prior distributions, $\mathcal{N}(\boldsymbol{w}_n|\boldsymbol{0},\boldsymbol{Q}_n(\boldsymbol{\theta}_w))$ and $\mathcal{N}(\boldsymbol{v}_n|\boldsymbol{0},\boldsymbol{R}_n(\boldsymbol{\theta}_v))$, $\forall n\in\mathbb{Z}^+$, are chosen for the $\boldsymbol{w}_n$ and $\boldsymbol{v}_n$ based on the Principle of Maximum (Information) Entropy \cite{jaynes1957information} under first and second moment constraints, where $\{\boldsymbol{w}_n\}_{n=1}^{N}$ and $\{\boldsymbol{v}_n\}_{n=1}^{N}$ are sequences of independent stochastic variables \cite{beck2010bayesian}. We add the uncertain parameters that specify these priors to the model parameters $\boldsymbol{\theta}_s$ and use $\boldsymbol{\theta}=[\boldsymbol{\theta}_s^T\,\,\boldsymbol{\theta}_w^T\,\,\boldsymbol{\theta}_v^T]^T$ to denote the uncertain parameter vector for the stochastic state-space model. If the initial state $\boldsymbol{x}_0$ is uncertain, we also add it to $\boldsymbol{\theta}$ and then choose a prior $p(\boldsymbol{\theta}|\mathcal{M}(\epsilon))$ for all of the model class parameters.

The defined stochastic state-space model defines a ``hidden'' Markov chain for the state time history $\{\boldsymbol{x}\}_{n=1}^{N}$ (which will also be denoted by the vector $\boldsymbol{x}_{1:N}=[\boldsymbol{x}_1^T,\,\dots,\,\boldsymbol{x}_N^T]^T\in\mathbb{R}^{NN_s}$) by implying a state transition PDF:
\begin{equation}\label{eq:eq8}
	\forall n\in\mathbb{Z}^+, \,\,p(\boldsymbol{x}_n|\boldsymbol{x}_{n-1},\boldsymbol{u}_{n-1},\boldsymbol{\theta},\mathcal{M}(\epsilon))=\mathcal{N}(\boldsymbol{x}_n|\boldsymbol{f}_n(\boldsymbol{x}_{n-1},\boldsymbol{u}_{n-1},\boldsymbol{\theta}),\boldsymbol{Q}_n(\boldsymbol{\theta}))
\end{equation}
along with a state-to-output PDF:
\begin{equation}\label{eq:eq9}
	\forall n\in\mathbb{Z}^+, \,\,p(\boldsymbol{y}_n|\boldsymbol{x}_n,\boldsymbol{u}_n,\boldsymbol{\theta},\mathcal{M}(\epsilon))=\mathcal{N}(\boldsymbol{y}_n|\boldsymbol{g}_n(\boldsymbol{x}_n,\boldsymbol{u}_n,\boldsymbol{\theta}),\boldsymbol{R}_n(\boldsymbol{\theta}))
\end{equation}

These, in turn, imply the following two probability models connecting the input, state and output discrete-time histories (which are readily sampled because each factor is Gaussian):
\begin{equation}\label{eq:eq10}
	p(\boldsymbol{x}_{1:N}|\boldsymbol{u}_{0:N},\boldsymbol{\theta},\mathcal{M}(\epsilon))=\prod_{n=1}^N p(\boldsymbol{x}_n|\boldsymbol{x}_{n-1},\boldsymbol{u}_{n-1},\boldsymbol{\theta},\mathcal{M}(\epsilon))
\end{equation}
\begin{equation}\label{eq:eq11}
	p(\boldsymbol{y}_{1:N}|\boldsymbol{x}_{1:N},\boldsymbol{u}_{0:N},\boldsymbol{\theta},\mathcal{M}(\epsilon))=\prod_{n=1}^N p(\boldsymbol{y}_n|\boldsymbol{x}_n,\boldsymbol{u}_n,\boldsymbol{\theta},\mathcal{M}(\epsilon))
\end{equation}

The stochastic input-output model (or forward model) for given parameter vector $\boldsymbol{\theta}$ is then:
\begin{equation}\label{eq:eq12}
	p(\boldsymbol{y}_{1:N}|\boldsymbol{u}_{0:N},\boldsymbol{\theta},\mathcal{M}(\epsilon))=\int p(\boldsymbol{y}_{1:N}|\boldsymbol{x}_{1:N},\boldsymbol{u}_{0:N},\boldsymbol{\theta},\mathcal{M}(\epsilon))p(\boldsymbol{x}_{1:N}|\boldsymbol{u}_{0:N},\boldsymbol{\theta},\mathcal{M}(\epsilon))\,d\boldsymbol{x}_{1:N}
\end{equation}
This high-dimensional integral usually cannot be done analytically. We will therefore structure the stochastic input-output model using a Bayesian hierarchical model to avoid the integration in (\ref{eq:eq12}), where $\boldsymbol{y}_{1:N}$ and $\boldsymbol{x}_{1:N}$ are treated in the same way as the model parameters $\boldsymbol{\theta}$.

This can be done by extending the stochastic model to predict the measured system output $\boldsymbol{z}_n$ at time $t_n$:
\begin{equation}\label{eq:eq13}
	\boldsymbol{z}_n=\boldsymbol{y}_n+\boldsymbol{e}_n=\boldsymbol{g}_n(\boldsymbol{x}_n,\boldsymbol{u}_n,\boldsymbol{\theta})+\boldsymbol{v}_n+\boldsymbol{e}_n
\end{equation}
where $\boldsymbol{e}_n$ denotes the uncertain measurement error at time $t_n$. Unlike $\{\boldsymbol{w}_n\}_{n=1}^N$ and $\{\boldsymbol{v}_n\}_{n=1}^N$, the stochastic sequence $\{\boldsymbol{e}_n\}_{n=1}^N$ will not necessarily be modeled as a set of independent stochastic variables; rather, we allow for dependence by specifying a joint PDF for $\boldsymbol{e}_{1:n}$ for any $n\in\mathbb{Z}^+$. 
A simple choice for the probability model for the measurement error $\boldsymbol{e}_{1:n}$ is a uniform PDF: 
\begin{equation}\label{eq:eq14}
	p(\boldsymbol{e}_{1:n}|\mathcal{M}(\epsilon))=\frac{\mathbb{I}_{\epsilon}(\boldsymbol{e}_{1:n})}{V_n(\epsilon)},\,\,\,\,\forall\boldsymbol{e}_{1:n}\in\mathbb{R}^{N_on}
\end{equation}
where $\mathbb{I}_{\epsilon}(\boldsymbol{e}_{1:n})$ is the indicator function for the set $S(\epsilon)=\{\boldsymbol{e}_{1:n}\in\mathbb{R}^{N_on}:\|\boldsymbol{e}_{1:n}\|\leq\epsilon\}$ for some vector norm (e.g., $\|.\|_\infty$ or $\|.\|_2$) on $\mathbb{R}^{N_o n}$, and $V_n(\epsilon)=\int_{\mathbb{R}^{N_on}}\mathbb{I}_{\epsilon} (\boldsymbol{e}_{1:n})\,d\boldsymbol{e}_{1:n}$ is the volume of region $S(\epsilon)$. This finally reveals that $\epsilon$ in $\mathcal{M}(\epsilon)$ is a scalar upper bound on the measurement errors that parameterizes the chosen prior PDF for $\boldsymbol{e}_{1:n}$, $\forall n\in\mathbb{Z}^+$. Thus, the predictive PDF for the observed system output (sensor output) $\boldsymbol{z}_{1:n}$, given the actual system output $\boldsymbol{y}_{1:n}$, is then given by:
\begin{equation}\label{eq:eq15}
	p(\boldsymbol{z}_{1:n}|\boldsymbol{y}_{1:n},\mathcal{M}(\epsilon))=p(\boldsymbol{e}_{1:n}|\mathcal{M}(\epsilon))\bigg|_{\boldsymbol{e}_{1:n}=\boldsymbol{z}_{1:n}-\boldsymbol{y}_{1:n}}=\left\{
	\begin{array}{ll}
	{V_n(\epsilon)}^{-1}\,\,&\mbox{if}\,\|\boldsymbol{z}_{1:n}-\boldsymbol{y}_{1:n}\|\leq\epsilon\\
	0\,&\mbox{otherwise}
	\end{array}
	\right.
\end{equation}

The specification of the hierarchical prior PDF:
\begin{equation}\label{eq:eq16}
		p(\boldsymbol{y}_{1:N},\boldsymbol{x}_{1:N},\boldsymbol{\theta}|\boldsymbol{u}_{0:N},\mathcal{M}(\epsilon))=p(\boldsymbol{y}_{1:N}|\boldsymbol{x}_{1:N},\boldsymbol{u}_{0:N},\boldsymbol{\theta},\mathcal{M}(\epsilon))p(\boldsymbol{x}_{1:N}|\boldsymbol{u}_{0:N},\boldsymbol{\theta},\mathcal{M}(\epsilon))\,p(\boldsymbol{\theta}|\mathcal{M}(\epsilon))
\end{equation}
completes the definition of the stochastic model class $\mathcal{M}(\epsilon)$. Thus, the PDFs in (\ref{eq:eq15}) and (\ref{eq:eq16}) define a hierarchical stochastic model class $\mathcal{M}(\epsilon)$ for each value of $\epsilon$ \cite{vakilzadeh2016approximate}. Here, $p(\boldsymbol{\theta}|\mathcal{M}(\epsilon))$ is assumed to be independent of the system input history $\boldsymbol{u}_{0:N}$.

\subsection{Bayesian model updating}\label{sec:BMU}
If measured system input and system output data:
\begin{equation*}
	\mathcal{D}_N=\{\hat{\boldsymbol{u}}_{0:N},\hat{\boldsymbol{z}}_{1:N}\}
\end{equation*}
are available from the dynamic system, then the predictive PDF in (\ref{eq:eq15}) with $n=N$ gives the likelihood function:
\begin{equation}\label{eq:eq17}
	p(\hat{\boldsymbol{z}}_{1:N}|\boldsymbol{y}_{1:N},\mathcal{M}(\epsilon))=\frac{\mathbb{I}_{\mathcal{D}(\epsilon)}(\boldsymbol{y}_{1:N})}{V_N(\epsilon)}
\end{equation}
with the indicator function defined over the set $\mathcal{D}(\epsilon)=\{\boldsymbol{y}_{1:N}\in\mathbb{R}^{NN_o}:\|\boldsymbol{y}_{1:N}-\hat{\boldsymbol{z}}_{1:N}\|<\epsilon\}$, where $\|.\|$ is some vector norm on $\mathbb{R}^{NN_o}$.

The posterior PDF for stochastic model class $\mathcal{M}(\epsilon)$ is then given by Bayes' Theorem:
\begin{equation}\label{eq:eq18}
	p(\boldsymbol{y}_{1:N},\boldsymbol{x}_{1:N},\boldsymbol{\theta}|\mathcal{D}_N,\mathcal{M}(\epsilon))=E(\epsilon)^{-1}\,\frac{\mathbb{I}_{\mathcal{D}(\epsilon)}(\boldsymbol{y}_{1:N})}{V_N(\epsilon)}\,p(\boldsymbol{y}_{1:N},\boldsymbol{x}_{1:N},\boldsymbol{\theta}|\hat{\boldsymbol{u}}_{0:N},\mathcal{M}(\epsilon))
\end{equation}
where the evidence for $\mathcal{M}(\epsilon)$ is then defined as:
\begin{equation}\label{eq:eq19}
	\begin{split}
	E(\epsilon)&=p(\hat{\boldsymbol{z}}_{1:N}|\hat{\boldsymbol{u}}_{0:N},\mathcal{M}(\epsilon))\\
	&=\int p(\hat{\boldsymbol{z}}_{1:N}|\boldsymbol{y}_{1:N},\epsilon)\,p(\boldsymbol{y}_{1:N},\boldsymbol{x}_{1:N},\boldsymbol{\theta}|\hat{\boldsymbol{u}}_{0:N},\mathcal{M}(\epsilon))\,d\boldsymbol{y}_{1:N}\,d\boldsymbol{x}_{1:N}\,d\boldsymbol{\theta}\\
	&=\int \frac{\mathbb{I}_{\mathcal{D}(\epsilon)}(\boldsymbol{y}_{1:N})}{V_N(\epsilon)}p(\boldsymbol{y}_{1:N},\boldsymbol{x}_{1:N},\boldsymbol{\theta}|\hat{\boldsymbol{u}}_{0:N},\mathcal{M}(\epsilon))\,d\boldsymbol{y}_{1:N}\,d\boldsymbol{x}_{1:N}\,d\boldsymbol{\theta}
	\end{split}
\end{equation}

The theory for the hierarchical model and its updating presented so far in Section \ref{sec:for} is general and valid for any $\epsilon>0$ deemed appropriate. For the application of ABC, we suppose that $\mathcal{M}(0)\equiv\mathcal{M}({\epsilon\rightarrow0})$ is actually the stochastic model class of interest. In this case, the output prediction errors $\boldsymbol{v}_n$ in (\ref{eq:eq7}) represent a combination of measurement errors and modeling errors with respect to the system output. For $\epsilon$ sufficiently small, the set $\mathcal{D}(\epsilon)$ of outputs $\boldsymbol{y}_{1:N}$ will converge to the observed output vector $\hat{\boldsymbol{z}}_{1:N}$ and the posterior PDF in (\ref{eq:eq18}) for stochastic model class $\mathcal{M}(\epsilon)$ will converge to the desired posterior distribution of the model parameters $p(\boldsymbol{\theta}|\mathcal{D}_N,\mathcal{M}(0))$ after marginalization. This can be shown in the following way:
\begin{equation}\label{eq:eq20}
	\begin{split}
	&p(\boldsymbol{\theta}|\mathcal{D}_N,\mathcal{M}(\epsilon))=\int p(\boldsymbol{y}_{1:N},\boldsymbol{x}_{1:N},\boldsymbol{\theta}|\mathcal{D}_N,\mathcal{M}(\epsilon))\,d\boldsymbol{y}_{1:N}\,d\boldsymbol{x}_{1:N}=\\
	&
	\frac{1}{E(\epsilon)}\int \frac{\mathbb{I}_{\mathcal{D}(\epsilon)}(\boldsymbol{y}_{1:N})}{V_N(\epsilon)}p(\boldsymbol{y}_{1:N}|\boldsymbol{x}_{1:N},\boldsymbol{\theta},\hat{\boldsymbol{u}}_{0:N},\mathcal{M}(\epsilon))p(\boldsymbol{x}_{1:N}|\boldsymbol{\theta},\hat{\boldsymbol{u}}_{0:N},\mathcal{M}(\epsilon))p(\boldsymbol{\theta}|\mathcal{M}(\epsilon))\,d\boldsymbol{y}_{1:N}\,d\boldsymbol{x}_{1:N}
	\end{split}
\end{equation}

Utilizing (\ref{eq:eq12}) to simplify the integrand in (\ref{eq:eq20}) and allowing $\epsilon\rightarrow0$ gives:
\begin{equation}\label{eq:eq22}
	\begin{split}
	p(\boldsymbol{\theta}|\mathcal{D}_N,\mathcal{M}(0))=& \frac{1}{E(0)}\int \delta_{\hat{\boldsymbol{z}}_{1:N}}(\boldsymbol{y}_{1:N})\,p(\boldsymbol{y}_{1:N}|\boldsymbol{\theta},\hat{\boldsymbol{u}}_{0:N},\mathcal{M}(0))\,p(\boldsymbol{\theta}|\mathcal{M}(0))\,d\boldsymbol{y}_{1:N}\\
	=&\,\frac{p(\hat{\boldsymbol{z}}_{1:N}|\boldsymbol{\theta},\hat{\boldsymbol{u}}_{0:N},\mathcal{M}(0))\,p(\boldsymbol{\theta}|\mathcal{M}(0))}{E(0)}
	\end{split}
\end{equation}
in which $E(0)\equiv E({\epsilon\rightarrow0})=\int p(\hat{\boldsymbol{z}}_{1:N}|\boldsymbol{\theta},\hat{\boldsymbol{u}}_{0:N},\mathcal{M}(0))\,p(\boldsymbol{\theta}|\mathcal{M}(0))\,d\boldsymbol{\theta}$, the evidence for $\mathcal{M}(0)$. However, if the tolerance $\epsilon$ is small, the acceptance rate is small so that the posterior distribution is estimated by only a few points unless the Algorithm \ref{algorithm:1} is run for a very long time. On the contrary, if the tolerance is too large, $(\epsilon\rightarrow\infty)$ then the samples come from the hierarchical prior $p(\boldsymbol{y}_{1:N},\boldsymbol{x}_{1:N},\boldsymbol{\theta}|\hat{\boldsymbol{u}}_{0:N},\mathcal{M}(\epsilon))$. Thus, a rational choice for $\epsilon$ should strike a balance between computability and accuracy. 

\begin{rmk}\label{rmk:1}
	Vakilzadeh et al. \cite{vakilzadeh2016approximate} showed that for the hierarchical stochastic model class $\mathcal{M}(\epsilon)$, the exact posterior PDF (\ref{eq:eq18}) using a uniformly-distributed uncertain measurement error in the output space is identical to the ABC posterior PDF given for no measurement error, see (\ref{eq:eq15}) and (\ref{eq:eq19}) in Vakilzadeh et al. \cite{vakilzadeh2016approximate}. Thus, ABC-SubSim that was originally developed by Chiachio et al. \cite{chiachio2014approximate} to draw samples from an ABC posterior PDF can be used to solve the exact Bayesian problem for the hierarchical stochastic model class $\mathcal{M}(\epsilon)$ given by (\ref{eq:eq15}) and (\ref{eq:eq16}).	
\end{rmk}

\subsubsection{Approximate Bayesian Computation by Subset Simulation (ABC-SubSim)}
For a good approximation of the posterior distribution for stochastic model class $\mathcal{M}(0)$, we want $\mathcal{D}(\epsilon)$ to be small neighborhood centered on the data vector $\hat{\boldsymbol{z}}_{1:N}$ in $\mathbb{R}^{NN_o}$. The probability $P(\boldsymbol{y}_{1:N}\in\mathcal{D}(\epsilon)|\epsilon)$ will then be small and so, on average, many candidate samples will be required to generate an acceptable sample having $\boldsymbol{y}_{1:N}\in\mathcal{D}(\epsilon)$ (on average, $1/P(\boldsymbol{y}_{1:N}\in\mathcal{D}(\epsilon)|\epsilon)$  candidate samples will be required). Henceforth, we use $P(\mathcal{D}(\epsilon))\equiv P(\boldsymbol{y}_{1:N}\in\mathcal{D}(\epsilon)|\epsilon)$ for a simpler notation. ABC-SubSim \cite{chiachio2014approximate} was originally developed to address this problem by exploiting the Subset Simulation method for efficient rare-event simulation \cite{au2001estimation}. The reader is referred to \cite{au2001estimation} and \cite{chiachio2014approximate} for a detailed explanation of how Subset Simulation and ABC-SubSim work.

The basic idea behind ABC-SubSim is to define the data-approximating region $\mathcal{D}(\epsilon)$ as the intersection of a set of nested decreasing data-approximating regions, $\mathcal{D}(\epsilon_j)$ of ``radius'' $\epsilon_j$, as defined above after (\ref{eq:eq17}), where $\epsilon_{j+1}<\epsilon_j$. The probability $P(\mathcal{D}(\epsilon))$ can be then estimated as a product of conditional probabilities:
\begin{equation}\label{eq:eq23}
	P(\mathcal{D}(\epsilon))=P(\mathcal{D}(\epsilon_1))\prod_{j=2}^{m}P(\mathcal{D}(\epsilon_j)|\mathcal{D}(\epsilon_{j-1}))
\end{equation}
The intermediate data-approximating regions are adaptively selected such that all conditional probabilities can be made large. Thus, ABC-SubSim replaces a problem involving rare-event simulation by a sequence of problems involving simulation of more frequent events.

The simulation algorithm starts by drawing $N_t$ independent and identically distributed samples $(\boldsymbol{y}_{1:N}^{i,(1)},\allowbreak\boldsymbol{x}_{1:N}^{i,(1)},\boldsymbol{\theta}^{i,(1)})$ from the hierarchical prior $p(\boldsymbol{y}_{1:N},\boldsymbol{x}_{1:N},\boldsymbol{\theta}|\boldsymbol{u}_{0:N},\mathcal{M})$. The corresponding metric value $\epsilon^{i,(1)}=\|y_{1:N}^{i,(1)}-\hat{\boldsymbol{z}}_{1:N}\|$ is then evaluated and samples are sorted in decreasing order of magnitude of their metric value so that $\epsilon^{1,(1)}\geq\epsilon^{2,(1)}\geq\dots\geq\epsilon^{N_t,(1)}$. Thus, probability $P(\mathcal{D}(\epsilon^{i,(1)}))$ corresponding to tolerance level $\epsilon^{i,(1)}$ can be approximated based on the samples by:
\begin{equation}\label{eq:eq24}
	P(\mathcal{D}(\epsilon^{i,(1)}))\approx\frac{1}{N_t}\sum_{i=1}^{N_t}\mathbb{I}_{\mathcal{D}(\epsilon^{i,(1)})}(\boldsymbol{y}_{1:N}^{i,(1)})=\frac{N_t-i}{N_t}=\hat{P}_{\epsilon^{i,(1)}}
\end{equation}
In ABC-SubSim, the initial tolerance level $\epsilon_1$ is chosen using (\ref{eq:eq24}) so that $\hat{P}_{\epsilon_1}=P_0$, an assigned probability whose value is best selected from the range $[0.1,\,0.3]$ (see Remark \ref{rmk:2}). 

For higher simulation levels $j\geq2$, sampling from the conditional PDF $(\boldsymbol{y}_{1:N},\boldsymbol{x}_{1:N},\boldsymbol{\theta}|\allowbreak\boldsymbol{y}_{1:N}\in\mathcal{D}(\epsilon_{j-1}),\hat{\boldsymbol{u}}_{0:N})$ can be achieved by means of a component-wise MCMC algorithm, called Modified Metropolis Algorithm (MMA) in \cite{au2001estimation}, at the expense of generating dependent samples. Using the MCMC samples $(\boldsymbol{y}_{1:N}^{i,(j)},\boldsymbol{x}_{1:N}^{i,(j)},\boldsymbol{\theta}^{i,(j)})$, the conditional probability can be estimated as:
\begin{equation}\label{eq:eq25}
	P(\mathcal{D}(\epsilon_j)|\mathcal{D}(\epsilon_{j-1}))\approx\frac{1}{N_t}\sum_{i=1}^{N_t}\mathbb{I}_{\mathcal{D}(\epsilon_j)}(\boldsymbol{y}_{1:N}^{i,(j)})=\hat{P}_{\epsilon_j|\epsilon_{j-1}}
\end{equation}
where $P(\mathcal{D}(\epsilon_j)|\mathcal{D}(\epsilon_{j-1}))\equiv P(\boldsymbol{y}_{1:N}\in \mathcal{D}(\epsilon_j)|\boldsymbol{y}_{1:N}\in\mathcal{D}(\epsilon_{j-1}))$ is the conditional probability at the $j$th simulation level. In ABC-SubSim, the intermediate tolerance levels $\epsilon_j$ ($j\geq2$) are adaptively determined as in Subset Simulation \cite{au2001estimation}, such that the sample estimate $\hat{P}_{\epsilon_j |\epsilon_{j-1}}$ of the conditional probabilities $P({\mathcal{D}(\epsilon_j)|\mathcal{D}(\epsilon_{j-1})})$ is equal to an assigned value $P_0$. To this end, we rearrange the samples generated by MCMC at the $j$th simulation level in decreasing order of the magnitude for their associated metric values $\{\epsilon^{i,(j)},\,i=1,\,\dots,\,N_t\}$. Then, the tolerance $\epsilon_j$ can be determined as the $100P_0$ percentile of the set of metric values $\epsilon^{i,(j)}, i=1,\,\dots,\,N_t$. For instance, we define $\epsilon_j=(\epsilon^{N_t (1-P_0 ),(j)}+\epsilon^{N_t (1-P_0 )+1,(j)})/2$.

Observe that the MCMC samples generated at the $j$th simulation level that fell in the data-approximation region $\mathcal{D}(\epsilon_j)$, i.e., samples corresponding to $\{\epsilon^{N_t (1-P_0 )+i,(j)},i=1,\,\dots,\,N_t P_0\}$, are distributed as $p(\boldsymbol{y}_{1:N},\allowbreak\boldsymbol{x}_{1:N},\boldsymbol{\theta}|\boldsymbol{y}_{1:N}\in \mathcal{D}(\epsilon_j),\hat{\boldsymbol{u}}_{0:N})$, and thus they provide $N_t P_0$ seeds in $\mathcal{D}(\epsilon_j)$. Thus, a Markov chain of length $(1-1/P_0)$ can be initiated from each of the seeds to populate $\mathcal{D}(\epsilon_j)$ with $N_t$ samples and it will be in its stationary state from the start, giving \textit{perfect sampling} (e.g., \cite{robert2013monte}).

Each of the sorted metric values $\epsilon^{i,(j)}$ gives a corresponding probability $P(\mathcal{D}(\epsilon^{i,(j)}))$, which can be approximated based on samples by:
\begin{equation}\label{eq:eq26}
	\begin{split}
		P(\mathcal{D}(\epsilon^{i,(j)}))&=P(\mathcal{D}(\epsilon_1))\,\bigg[\prod_{k=2}^{j-1}P({\mathcal{D}(\epsilon_k)|\mathcal{D}(\epsilon_{k-1})})\bigg]\,P(\mathcal{D}(\epsilon^{i,(j)})|\mathcal{D}(\epsilon_{j-1}))\\
		&\approx P_0^{j-1}\frac{N_t-i}{N_t}=\hat{P}_{\epsilon^{i,(j)}}
	\end{split}
\end{equation}

The algorithm proceeds in this way until $\epsilon_m$ becomes smaller than an appropriate final tolerance level $\epsilon$.
\begin{rmk}\label{rmk:2}
	To run the ABC-SubSim Algorithm, one needs to specify appropriate values for: (\textit{i}) the number of samples per simulation level $N_t$, and (\textit{ii}) the conditional probability $P_0$ of each stage of ABC-SubSim. Choosing a large value for $P_0$ increases the number of simulation levels required to achieve a specified tolerance level $\epsilon$ for a fixed $N_t$. Thus, the higher $P_0$ is, the higher is the computational burden of the algorithm. On the other hand, a choice of a small value for $P_0$ decreases the quality of the posterior approximation. Recently, Zuev et al. \cite{zuev2012bayesian} implemented a rigorous sensitivity analysis of Subset Simulation and reported that an optimal choice for $P_0$ lies in the range $[0.1,0.3]$. Furthermore, for convenience $P_0 N_t$ and $1/P_0$ are selected as positive integers.	
\end{rmk}
\begin{rmk}\label{rmk:3}
	(\ref{eq:eq24}) and (\ref{eq:eq26}) produce an estimate of the probability $P(\boldsymbol{y}_{1:N}\in\mathcal{D}(\epsilon))$ as a function of tolerance level $\epsilon$, covering the large probabilities to small probability regimes (e.g., see Figure \ref{fig:fig5}). This in turn means that the calculation of the evidence of a candidate model class for different tolerance levels is a simple by-product of the ABC-SubSim algorithm. In the next section, we will show that this property makes ABC-SubSim an effective algorithm for Bayesian model selection.	
\end{rmk}
\begin{rmk}\label{rmk:4}
	In \cite{vakilzadeh2016approximate}, a modification of ABC-SubSim, called self-regulating ABC-SubSim, has been proposed. The key idea behind this method is to enhance efficient exploration of the posterior distribution over the parameter space. A way to achieve this goal is to learn the proposal variance for the MMA algorithm in each simulation level on-the-fly in order to coerce the mean acceptance probability for a candidate sample to be close to a desired target value. Another key benefit of incorporating the self-regulating algorithm in ABC-SubSim is that it gives a heuristic rule to automatically determine the number of simulation levels m as follows: \textit{stop the algorithm when the average acceptance rate drops significantly}. This recently proposed variant of ABC-SubSim is used in this study to draw samples from the posterior PDF $p(\boldsymbol{y}_{1:N},\boldsymbol{y}_{1:N},\boldsymbol{\theta}|\mathcal{D}_N,\mathcal{M}(\epsilon))$ of stochastic model class $\mathcal{M}(\epsilon)$. 	
\end{rmk}

\subsection{Bayesian model class assessment}
Bayesian model class selection provides a rigorous framework to compare the performance of a set of candidate model classes in describing the experimental data. As exposed by Marin et al. \cite{marin2012approximate}, there are well-known limitations of the ABC approach to the Bayesian model selection problem, mainly due to lack of a sufficient summary statistics that work across models. However, in Section \ref{sec:BMU} we showed that formulating the standard ABC posterior distribution as the exact posterior PDF (\ref{eq:eq18}) for the hierarchical state-space model class allows us to independently estimate the evidence $E(\epsilon)$ for each alternative candidate model $\mathcal{M}(\epsilon)$ as the normalizing constant associated with the exact posterior PDF (\ref{eq:eq18}). Didelot et al. \cite{didelot2011likelihood} demonstrated that under mild continuity conditions, this normalizing constant converges to the true model evidence $E(0)$ as $\epsilon\rightarrow0$. In this section, our objective is to show that calculation of the evidence $E(\epsilon)$ is a simple by-product of the ABC-SubSim algorithm.

Consider a set $\boldsymbol{M} \equiv \{ \mathcal{M}_1(\epsilon_{\scriptscriptstyle\mathcal{M}_1}),\,\mathcal{M}_2(\epsilon_{\scriptscriptstyle\mathcal{M}_2}),\,\dots,\,\mathcal{M}_{L} (\epsilon_{\scriptscriptstyle\mathcal{M}_{L}})\}$ of $L$ Bayesian hierarchical model classes for representing a system. In Bayesian model selection, models in $\boldsymbol{M}$ are ranked based on their probabilities conditioned on the data $\mathcal{D}_N$ that is given by Bayes' Theorem:
\begin{equation}\label{eq:eq27}
	P(\mathcal{M}_j(\epsilon_{\scriptscriptstyle\mathcal{M}_j})|\mathcal{D}_N)=\frac{p(\hat{\boldsymbol{z}}_{1:N}|\hat{\boldsymbol{u}}_{0:N},\mathcal{M}_j(\epsilon_{\scriptscriptstyle\mathcal{M}_j}))P(\mathcal{M}_j(\epsilon_{\scriptscriptstyle\mathcal{M}_j})|\boldsymbol{M})}{\sum_{l=1}^{L}p(\hat{\boldsymbol{z}}_{1:N}|\hat{\boldsymbol{u}}_{0:N},\mathcal{M}_j(\epsilon_{\scriptscriptstyle\mathcal{M}_j}))P(\mathcal{M}_j(\epsilon_{\scriptscriptstyle\mathcal{M}_j})|\boldsymbol{M})}
\end{equation}
where $P(\mathcal{M}_j(\epsilon_{\scriptscriptstyle\mathcal{M}_j})|\boldsymbol{M})$ denotes the prior probability of $\mathcal{M}_j(\epsilon_{\scriptscriptstyle\mathcal{M}_j})$ that indicates the modeler's belief about the initial relative plausibility of $\mathcal{M}_j(\epsilon_{\scriptscriptstyle\mathcal{M}_j})$ within the set $\boldsymbol{M}$. The factor $p(\hat{\boldsymbol{z}}_{1:N}|\hat{\boldsymbol{u}}_{0:N},\mathcal{M}_j(\epsilon_{\scriptscriptstyle\mathcal{M}_j}))$, which is the evidence (or marginal likelihood) for $\mathcal{M}_j(\epsilon_{\scriptscriptstyle\mathcal{M}_j})$, indicates the probability of data $\mathcal{D}_N$ according to $\mathcal{M}_j(\epsilon_{\scriptscriptstyle\mathcal{M}_j})$.

For the specific choice of Bayesian hierarchical model class, the evidence can be estimated by (\ref{eq:eq19}). However, its calculation requires the evaluation of a high-dimensional integral which is the computationally challenging step in Bayesian model selection, especially as $\epsilon\rightarrow0$. ABC-SubSim provides a straightforward approximation for it via the conditional probabilities involved in the Subset Simulation. Indeed, the last integral in (\ref{eq:eq19}) is the probability $P(\boldsymbol{y}_{1:N}\in\mathcal{D}(\epsilon_{\scriptscriptstyle\mathcal{M}_j})|\mathcal{M}_j)$ that $\boldsymbol{y}_{1:N}$ belongs to $\mathcal{D}(\epsilon_{\scriptscriptstyle\mathcal{M}_j})=\{\boldsymbol{y}_{1:N}\in\mathbb{R}^{NN_o}: \|\boldsymbol{y}_{1:N}-\hat{\boldsymbol{z}}_{1:N}\|\leq\epsilon_{\scriptscriptstyle\mathcal{M}_j}\}$. This probability can be readily estimated as a by-product of ABC-SubSim by using (\ref{eq:eq24}) and (\ref{eq:eq26}). Thus, for a particular tolerance level $\epsilon_{\scriptscriptstyle\mathcal{M}_j}$ and model class $\mathcal{M}_j(\epsilon_{\scriptscriptstyle\mathcal{M}_j})$, the evidence is estimated by:
\begin{equation}\label{eq:eq28}
\hat{E}_{\scriptscriptstyle\mathcal{M}_j}=\frac{P(\boldsymbol{y}_{1:N}\in\mathcal{D}(\epsilon_{\scriptscriptstyle\mathcal{M}_j})|\mathcal{M}_j)}{V_N(\epsilon_{\scriptscriptstyle\mathcal{M}_j})}=\frac{1}{V_N(\epsilon_{\scriptscriptstyle\mathcal{M}_j})}P_0^{i-1}\,P_i
\end{equation}
where $i$ would be such that $\epsilon_i\leq\epsilon_{\scriptscriptstyle\mathcal{M}_j}<\epsilon_{i-1}$, in which the intermediate ``radii'' $\epsilon_i$'s are automatically chosen by ABC-SubSim, and $P_i$ is the fraction of samples generated in $\mathcal{D}(\epsilon_{i-1})$ that lie in $\mathcal{D}(\epsilon_{\scriptscriptstyle\mathcal{M}_j})$. Here, $V_N(\epsilon)$ is the volume of the ball centered at $\hat{\boldsymbol{z}}_{1:N}$, with radius $\epsilon$ and norm $\|.\|$. If $\mathcal{D}(\epsilon)$ is defined by using $\|.\|_\infty$ on $\mathbb{R}^{NN_o}$, then $V_N(\epsilon)=(2\epsilon)^{NN_o}$ and if is defined by the Euclidean norm, then
$V_N(\epsilon)=\pi^{NN_o/2}/\Gamma(NN_o/2+1)\epsilon^{NN_o}$. It is worth noting that $V_N(\epsilon_{\scriptscriptstyle\mathcal{M}_j})$ is not needed for posterior model class assessment if we choose the same tolerance level in ABC for each of the $L$ candidate model classes.
Wilkinson \cite{wilkinson2008bayesian,wilkinson2013approximate} showed that a standard ABC posterior gives an exact posterior distribution for a new model under the assumption that the summary statistics are corrupted with a uniform additive error term. However, formulating standard ABC based on summary statistics hinders the independent approximation of evidence for each candidate model \cite{barthelme2014expectation}. Here, the estimate of the model evidence in (\ref{eq:eq28}) is a result of formulating a dynamic problem in terms of a general hierarchical stochastic state-space model where the likelihood function $p(\hat{\boldsymbol{z}}_{1:N}|\boldsymbol{y}_{1:N},\mathcal{M}(\epsilon))$ is expressed using the entire data $\hat{\boldsymbol{z}}_{1:N}$ and ABC-SubSim readily produces an unbiased approximation of the evidence.

\subsection{Treatment of uncertain prediction error model parameters }
In reality the parameters specifying the covariance matrix of the uncertain prediction errors are unknown and they should be learned from the data $\mathcal{D}_N$. However, Vakilzadeh et al. \cite{vakilzadeh2016approximate} found that shrinking the data-approximation region, $\mathcal{D}(\epsilon\rightarrow0)$, drives the covariance matrix of the prediction errors to zero. This is due to the fact that the formulation of the stochastic state-space model has built-in prediction errors such that a sample of the model prediction of the system output includes a realization of these error signals. For high-dimensional data vectors, the probability of drawing sequences of random numbers $\{\boldsymbol{w}_n \}_{n=1}^N$ and $\{\boldsymbol{v}_n \}_{n=1}^N$ such that the measured data exactly matches the simulated system output is essentially zero, even if the measured data is synthetic and generated by adding realizations of the prediction-error signals when calculating the output from the chosen model for system identification. Therefore, one can conclude that not only are the parameters specifying the uncertain prediction errors unidentifiable based on the likelihood function $p(\hat{\boldsymbol{z}}_{1:N}|\boldsymbol{y}_{1:N},\mathcal{M}(\epsilon))$ constructed from the entire data, but also that when $\epsilon\rightarrow0$, the stochastic input-to-output model (\ref{eq:eq12}) reduces to the underlying deterministic model. 

To get around this problem, we treat the parameters $\boldsymbol{\theta}_w$ and $\boldsymbol{\theta}_v$ specifying the state and output uncertain errors as nuisance parameters and eliminate them from the analysis to achieve a posterior distribution for the structural model parameters $\boldsymbol{\theta}_s$ alone. Vakilzadeh et al. \cite{vakilzadeh2016approximate} defined appropriate conjugate priors for the $\boldsymbol{\theta}_w$ and $\boldsymbol{\theta}_v$ and then integrated them out from the posterior distribution (\ref{eq:eq18}) and worked with the resulting marginal posterior distribution for $\boldsymbol{\theta}_s$. Although the marginalization technique automatically incorporates the posterior uncertainty induced by the nuisance parameters \cite{berger1999integrated}, it relies on choosing appropriate prior distributions for the nuisance parameters. Here, we present an alternative approach that avoids having to choose these prior distributions by using Laplace's method to replace the marginal posterior distribution of $\boldsymbol{\theta}_s$ by its asymptotic approximation \cite{tierney1986accurate}, which turns out to be insensitive to the prior distribution adopted for the nuisance parameters. However, the use of this approximate method to solve the Bayesian problem for the hierarchical model class $\mathcal{M}(\epsilon)$ given by (\ref{eq:eq15}) and (\ref{eq:eq16}) is restricted to the cases where the state prediction error $\boldsymbol{w}_n$ is suppressed in (\ref{eq:eq7}) so that the hierarchical prior PDF in (\ref{eq:eq16}) is replaced by:
\begin{equation}\label{eq:eq30}
	p(\boldsymbol{y}_{1:N},\boldsymbol{\theta}|\boldsymbol{u}_{0:N},\mathcal{M}(\epsilon))=p(\boldsymbol{y}_{1:N}|\boldsymbol{u}_{0:N},\boldsymbol{\theta},\mathcal{M}(\epsilon))\,p(\boldsymbol{\theta}|\mathcal{M}(\epsilon))
\end{equation}
Actually, it is quite common in Bayesian inference for dynamic systems that the uncertainty in the system output is modeled by only an output prediction error $\boldsymbol{v}_n$. Furthermore, we model the covariance matrix for $\boldsymbol{v}_n$ as a time-invariant isotropic diagonal matrix $\boldsymbol{R}_n(\boldsymbol{\theta}_v )=\sigma^2 \boldsymbol{I}_{N_o}$ giving the scalar nuisance parameter $\theta_v=\sigma^2$. Throughout this section, we drop the conditioning on the model class $\mathcal{M}(\epsilon)$ since the formulation is valid for any choice of model class.

For reasons which will become clearer later, we temporarily neglect the measurement error in (\ref{eq:eq13}) and assume that the predictive PDF for the sensor output can be expressed by (\ref{eq:eq12}). For data $\mathcal{D}_N$, this gives the more common likelihood function $p(\hat{\boldsymbol{z}}_{1:N}|\boldsymbol{u}_{0:N},\boldsymbol{\theta})$ where $\boldsymbol{v}_n$ in (\ref{eq:eq7}) is often taken as a combination of measurement and modeling errors. The marginal likelihood function of the model parameters $\boldsymbol{\theta}_s$ is now:
\begin{equation}\label{eq:eq31}
p(\hat{\boldsymbol{z}}_{1:N}|\hat{\boldsymbol{u}}_{0:N},\boldsymbol{\theta}_s)=\int p(\hat{\boldsymbol{z}}_{1:N}|\hat{\boldsymbol{u}}_{0:N},\boldsymbol{\theta}_s,\theta_v)\,p(\theta_v)\,d\theta_v
\end{equation}

Our objective here is to develop an asymptotic approximation for this integral. To this end, suppose that $\hat{\theta}_v (\boldsymbol{\theta}_s)$ maximizes the likelihood function $p(\hat{\boldsymbol{z}}_{1:N}|\boldsymbol{u}_{0:N},\boldsymbol{\theta}_s,\theta_v)$ for a fixed value of the model parameters $\boldsymbol{\theta}_s$. By expanding $\ln p(\hat{\boldsymbol{z}}_{1:N}|\hat{\boldsymbol{u}}_{0:N},\boldsymbol{\theta}_s,\theta_v)$ in a second-order Taylor series about $\hat{\theta}_v (\boldsymbol{\theta}_s)$, one obtains the following local approximation \cite{beck1998updating}:
\begin{equation}\label{eq:eq32}
	p(\hat{\boldsymbol{z}}_{1:N}|\hat{\boldsymbol{u}}_{0:N},\boldsymbol{\theta}_s,\theta_v)=p(\hat{\boldsymbol{z}}_{1:N}|\hat{\boldsymbol{u}}_{0:N},\boldsymbol{\theta}_s,\hat{\theta}_v(\boldsymbol{\theta}_s))\exp\bigg(-\frac{1}{2}H(\hat{\theta}_v(\boldsymbol{\theta}_s))[\theta_v-\hat{\theta}_v(\boldsymbol{\theta}_s)]^2\bigg)
\end{equation}
where the Hessian $H(\hat{\theta}_v(\boldsymbol{\theta}_s))$ is given by:
\begin{equation}\label{eq:eq33}
	H(\hat{\theta}_v(\boldsymbol{\theta}_s))=-\frac{\partial^2\ln p(\hat{\boldsymbol{z}}_{1:N}|\hat{\boldsymbol{u}}_{0:N},\boldsymbol{\theta})}{\partial\theta_v^2}\Bigg|_{\theta_v=\hat{\theta}_v(\boldsymbol{\theta}_s)}
\end{equation}
and the MLE (maximum likelihood estimate) of the nuisance parameter $\theta_v$ for a fixed value of the model parameters $\boldsymbol{\theta}_s$ is equal to the mean prediction error value \cite{beck1998updating}:
\begin{equation}\label{eq:eq34}
	\hat{\theta}_v(\boldsymbol{\theta}_s)=\frac{1}{NN_o}\Big\|\hat{\boldsymbol{z}}_{1:N}-\boldsymbol{g}_{1:N}(\boldsymbol{\theta}_s)\Big\|_2^2
\end{equation}
where $\boldsymbol{g}_{1:N}=[\boldsymbol{g}_1^T,\,\boldsymbol{g}_2^T,\,\dots,\,\boldsymbol{g}_N^T]^T$. Here $\boldsymbol{g}_n(\boldsymbol{\theta}_s)$ denotes $\boldsymbol{g}_n(\hat{\boldsymbol{x}}_n,\,\boldsymbol{u}_n,\,\boldsymbol{\theta}_s)$ from (\ref{eq:eq6}) where $\hat{\boldsymbol{x}}_n$ is the deterministic solution of $\boldsymbol{x}_n=\boldsymbol{f}_n(\boldsymbol{x}_{n-1},\,\hat{\boldsymbol{u}}_{n-1},\,\boldsymbol{\theta}_s)$ from (\ref{eq:eq6}). For a large number $N$ of sampling times, and given structural model parameters $\boldsymbol{\theta}_s$, $p(\hat{\boldsymbol{z}}_{1:N}|\hat{\boldsymbol{u}}_{0:N},\boldsymbol{\theta}_s,\theta_v)$ will be very peaked at the optimal parameter $\hat{\theta}_v(\boldsymbol{\theta}_s)$. Therefore, Laplace's method for asymptotic approximation can be applied to the integral in (\ref{eq:eq31}) to obtain the following approximation for the marginal likelihood distribution, which is essentially derived by substituting (\ref{eq:eq32}) into (\ref{eq:eq31}) \cite{beck1998updating,sweeting1987discussion}:
\begin{equation}\label{eq:eq35}
	p(\hat{\boldsymbol{z}}_{1:N}|\hat{\boldsymbol{u}}_{0:N},\boldsymbol{\theta}_s)=\sqrt{2\pi}{H\Big(\hat{\theta}_v(\boldsymbol{\theta}_s)\Big)}^{-\frac{1}{2}}\,p(\hat{\theta}_v(\boldsymbol{\theta}_s))p\bigg(\hat{\boldsymbol{z}}_{1:N}|\hat{\boldsymbol{u}}_{0:N},\boldsymbol{\theta}_s,\hat{\theta}_v(\boldsymbol{\theta}_s)\bigg)[1+\mathcal{O}(N^{-1})]
\end{equation}
\begin{rmk}\label{rmk:5}
	The structure of the Hessian matrix of $\ln p(\hat{\boldsymbol{z}}_{1:N}|\hat{\boldsymbol{u}}_{0:N},\boldsymbol{\theta})$ regarding the uncertain parameter vector $\boldsymbol{\theta}=[\boldsymbol{\theta}_s^T\, \theta_v]^T$  is block diagonal with one block being an $N_p\times N_p$ matrix corresponding to the structural model parameters $\boldsymbol{\theta}_s$ and the other block being simply the scalar $H(\hat{\theta}_v(\boldsymbol{\theta}_s))=\frac{1}{2}NN_o\hat{\theta}_v(\boldsymbol{\theta}_s)^{-2}$ corresponding to the scalar nuisance parameter $\theta_v$ \cite{beck1998updating}. In this setting, Cox and Reid \cite{cox1987parameter} showed that the MLE $\hat{\theta}_v(\boldsymbol{\theta}_s)$ of the nuisance parameter can be treated approximately as a constant $\hat{\theta}_v$ in $p(\hat{\theta}_v(\boldsymbol{\theta}_s))$ and $H(\hat{\theta}_v(\boldsymbol{\theta}_s))$. We will show next that using this feature, the posterior distribution of the model parameters $\boldsymbol{\theta}_s$ is approximately free from the prior adopted for the nuisance parameter $\theta_v$ \cite{tierney1986accurate}.	
\end{rmk}

By substituting the asymptotic approximation of the likelihood function (\ref{eq:eq35}) into Bayes' Theorem, the posterior PDF can be approximated by:
\begin{equation}\label{eq:eq36}
p(\boldsymbol{\theta}_s|\mathcal{D}_N)\cong\frac{\sqrt{2\pi}{H(\hat{\theta}_v)}^{-\frac{1}{2}}p(\hat{\theta}_v)\,p\big(\hat{\boldsymbol{z}}_{1:N}|\hat{\boldsymbol{u}}_{0:N},\boldsymbol{\theta}_s,\hat{\theta}_v(\boldsymbol{\theta}_s)\big)\,p(\boldsymbol{\theta}_s)}{\int \sqrt{2\pi}{H(\hat{\theta}_v)}^{-\frac{1}{2}}p(\hat{\theta}_v)\,p\big(\hat{\boldsymbol{z}}_{1:N}|\hat{\boldsymbol{u}}_{0:N},\boldsymbol{\theta}_s,\hat{\theta}_v(\boldsymbol{\theta}_s)\big)\,p(\boldsymbol{\theta}_s)\,d\boldsymbol{\theta}_{s}}
\end{equation}
where $\cong$ denotes an approximation to $\mathcal{O}(N^{-1})$. Using Remark \ref{rmk:5}, the asymptotic approximation for the marginal posterior distribution of the structural model parameters $\boldsymbol{\theta}_s$ can be rewritten as:
\begin{equation}\label{eq:eq37}
p(\boldsymbol{\theta}_s|\mathcal{D}_N)\cong\frac{p\big(\hat{\boldsymbol{z}}_{1:N}|\hat{\boldsymbol{u}}_{0:N},\boldsymbol{\theta}_s,\hat{\theta}_v(\boldsymbol{\theta}_s)\big)\,p(\boldsymbol{\theta}_s)}{\int p\big(\hat{\boldsymbol{z}}_{1:N}|\hat{\boldsymbol{u}}_{0:N},\boldsymbol{\theta}_s,\hat{\theta}_v(\boldsymbol{\theta}_s)\big)\,p(\boldsymbol{\theta}_s)\,d\boldsymbol{\theta}_{s}}\propto p\big(\hat{\boldsymbol{z}}_{1:N}|\hat{\boldsymbol{u}}_{0:N},\boldsymbol{\theta}_s,\hat{\theta}_v(\boldsymbol{\theta}_s)\big)\,p(\boldsymbol{\theta}_s)
\end{equation}
which shows that the prior distribution of the nuisance parameter cancels out. Following the same line of thought as for ABC methods (see Introduction), the approximation of the marginal posterior distribution (\ref{eq:eq37}) can be augmented to:
\begin{equation}\label{eq:eq38}
p(\boldsymbol{\theta}_s|\mathcal{D}_N)\propto p(\hat{\boldsymbol{z}}_{1:N}|\boldsymbol{y}_{1:N},\epsilon)\,p\big(\boldsymbol{y}_{1:N}|\hat{\boldsymbol{u}}_{0:N},\boldsymbol{\theta}_s,\hat{\theta}_v(\boldsymbol{\theta}_s)\big)\,p(\boldsymbol{\theta}_s)
\end{equation}

By defining $p(\hat{\boldsymbol{z}}_{1:N}|\boldsymbol{y}_{1:N},\epsilon)=p(\boldsymbol{e}_{1:N}|\epsilon)$ as the uniform probability distribution given in (\ref{eq:eq17}), the ABC approximate marginal posterior distribution (\ref{eq:eq38}) gives the exact posterior PDF for the hierarchical state-space model class with a uniform measurement error for predictions of the sensor output and no state prediction errors $\boldsymbol{w}_n$ \cite{vakilzadeh2016approximate}. 

\section{Illustrative examples for Bayesian system identification}
The utilization of Bayesian framework for system identification has received increasing attention in recent years (e.g., \cite{beck2010bayesian,beck2004model,cheung2010calculation,muto2008bayesian,beck1998updating,ching2007TMCMC,cheung2009bayesian,angelikopoulos2015x,au2016fundamental,beck2002bayesian,goller2009robust,green2015bayesian,jensen2013use,madireddy2015bayesian,straub2014bayesian,worden2012parameter,ching2006bayesian,ching2006bayesian2,ching2006structural,yuen2003updating,yuen2010recent,papadimitriou2001updating}). Here, we study two numerical examples from Bayesian system identification literature to demonstrate the application of the ABC-SubSim algorithm to model class updating and selection, with a special focus on the estimation of the model class evidence for different values of the tolerance level $\epsilon$. The first example, which is a single degree-of-freedom bilinear oscillator, compares the performance of ABC-SubSim for Bayesian model class selection with the method presented in Beck and Yuen \cite{beck2004model}, which provides an asymptotic approximation for the model evidence in the presence of large amounts of dynamic data. The second example, which is a three degree-of-freedom nonlinear structure, demonstrates the applicability of ABC-SubSim to perform Bayesian model class selection for a class of Masing hysteretic models and compares its behavior with the TMCMC algorithm \cite{ching2007TMCMC}. Both examples use input and output data that is artificially generated by subjecting the dynamic model to recorded seismic excitations. For ABC-SubSim, the self-regulating algorithm presented in \cite{vakilzadeh2016approximate} is used for both examples with the number of samples in each level fixed to $N_t=2000$ and with the adaptation probability $P_a=0.1$ and the optimal acceptance rate $\alpha^*=0.5$. However, the conditional probability for each level of ABC-SubSim is set differently for the first and second examples as $P_0=0.2$ and $P_0=0.1$, respectively. For both examples, our objective is to draw samples from the posterior $p(\boldsymbol{y}_{1:N},\boldsymbol{x}_{1:N},\boldsymbol{\theta}|\mathcal{D}_N,\epsilon)$ where $\boldsymbol{y}_{1:N}$ is constrained to lie in a small neighborhood, $\mathcal{D}(\epsilon)$, of the data vector defined by:
\begin{equation}\label{eq:eq42}
	\mathcal{D}(\epsilon)=\{\boldsymbol{y}_{1:N}\in\mathbb{R}^{NN_o}: \big\|\boldsymbol{y}_{1:N}-\hat{\boldsymbol{z}}_{1:N}\big\|_2\leq\epsilon\}
\end{equation}
The model evidence is then estimated as a by-product of ABC-SubSim.

\subsection*{Example 1: Single degree-of-freedom bilinear hysteretic oscillator under seismic excitation}
\begin{figure}[htpb!]
	\centering
	\includegraphics[width=.55\textwidth]{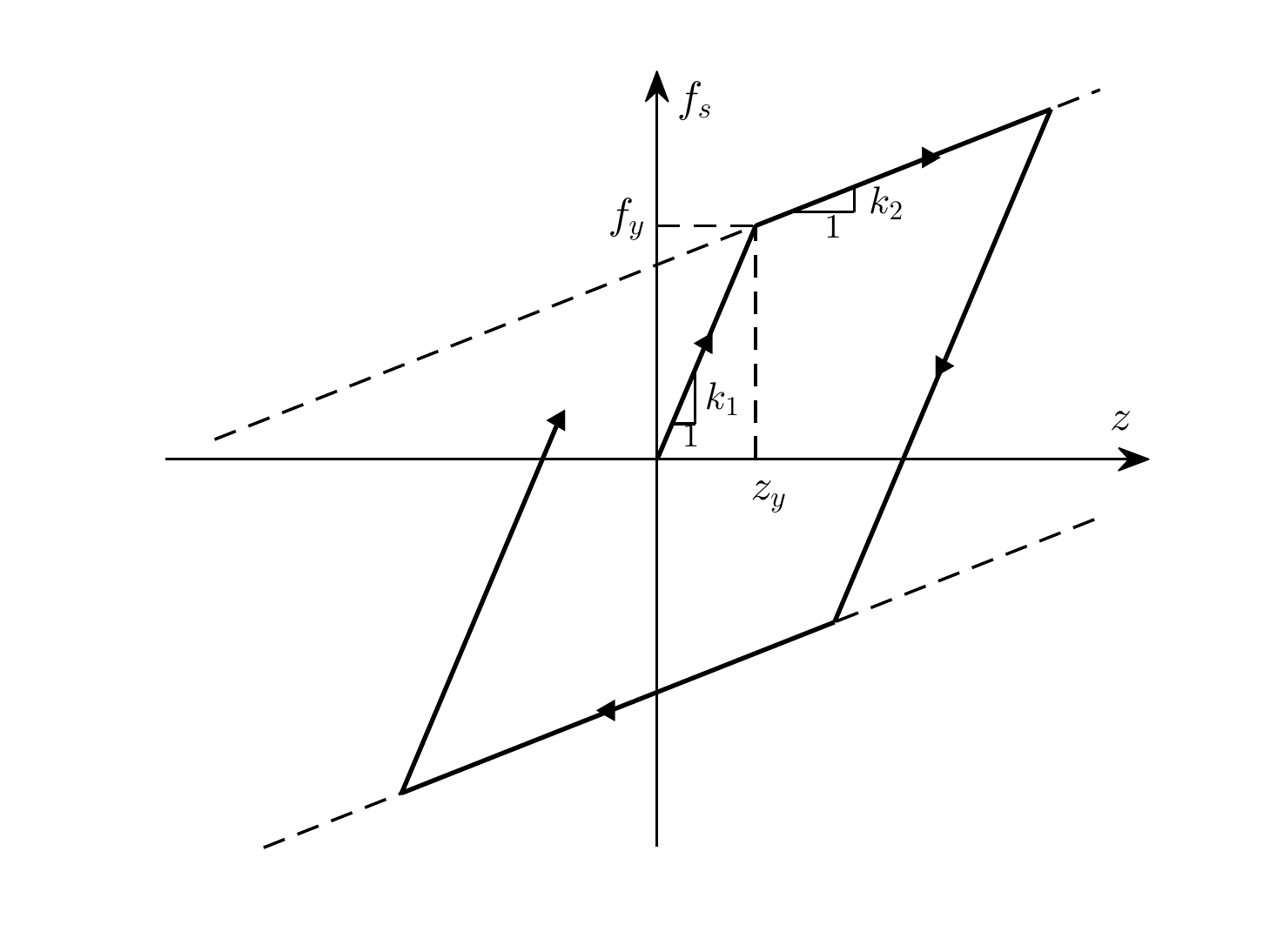}
	\vspace{-10pt}
	\caption{\emph{The hysteresis loop for the bilinear oscillator (Example 1).}}
	\label{fig:fig1}
\end{figure}

This example follows the spirit of the first example used in Beck and Yuen \cite{beck2004model} where response measurements are used to select the most plausible model class for different levels of excitation. In their paper, the model evidence is approximated by using Laplace's method for asymptotic approximation while we use ABC-SubSim here. The system is a single degree-of-freedom bilinear hysteretic oscillator with linear viscous damping. The equation of motion for this oscillator subject to ground acceleration can be represented by:
\begin{equation}\label{eq:eq43}
	m\ddot{z}(t)+c\dot{z}(t)+f_h(z;k_1,k_2,z_y)=-mu(t)
\end{equation}
where $z(t)\in\mathbb{R}$ is the horizontal displacement vector relative to ground; $u(t)\in\mathbb{R}$ is the horizontal seismic ground acceleration; $m$, $c\in\mathbb{R}$ denote the mass and linear viscous damping, respectively; $f_h\in\mathbb{R}$ denotes the hysteretic restoring force; $k_1\in\mathbb{R}$ is the elastic stiffness; $k_2\in\mathbb{R}$ is the post-yield stiffness; and $z_y\in\mathbb{R}$ is the yield displacement. The relationship between the restoring force and the displacement is shown in Figure \ref{fig:fig1}. 

The oscillator is assumed to have known mass $m=1$ kg. In general, $\boldsymbol{\theta}_s=[k_1\,\,k_2\,\,z_y\,\,c]^T$ forms the uncertain structural model parameter vector for the bilinear hysteresis oscillator where the actual values are defined as $k_1=1.0$ N/m, $k_2=0.1$ N/m, $c=0.02$ N.s/m and $z_y=2.0$ cm. Samples of the response time history $\boldsymbol{y}_{1:N}$ for given values of the uncertain parameters and ground acceleration time history are simulated using the function `ODE45' in Matlab.

\begin{figure}[htbp!]
	\begin{subfigure}[b]{0.5\columnwidth}
		\caption{}
		\vspace*{-7pt}
		\centering
		\setlength\fheight{4cm}
		\setlength\fwidth{6cm}
		\includegraphics[scale=0.45]{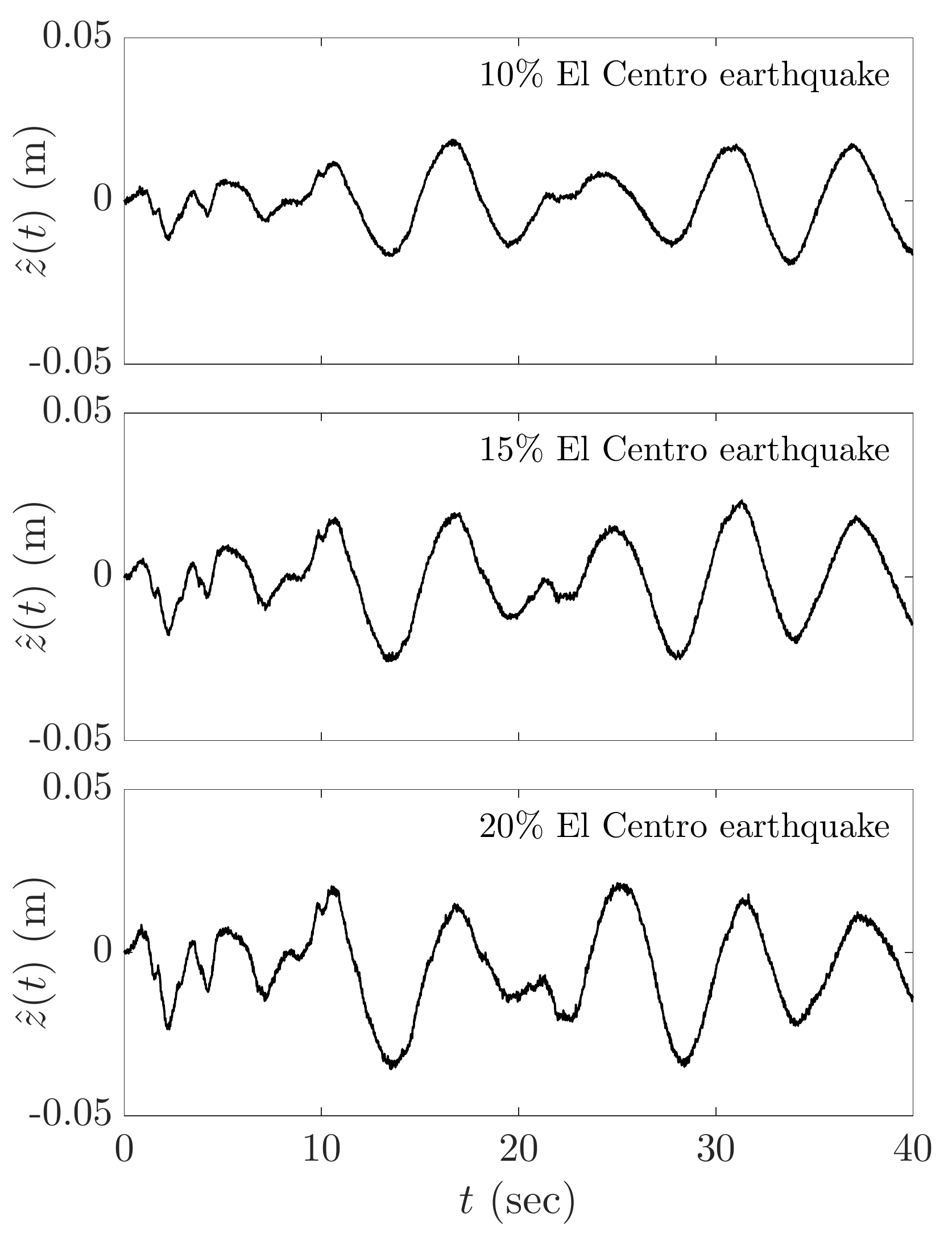}
		\label{fig:fig2a}
	\end{subfigure}
	\begin{subfigure}[b]{0.5\columnwidth}
		\caption{}
		\vspace*{-7pt}
		\centering
		\setlength\fheight{4cm}
		\setlength\fwidth{6cm}
		\includegraphics[scale=0.45]{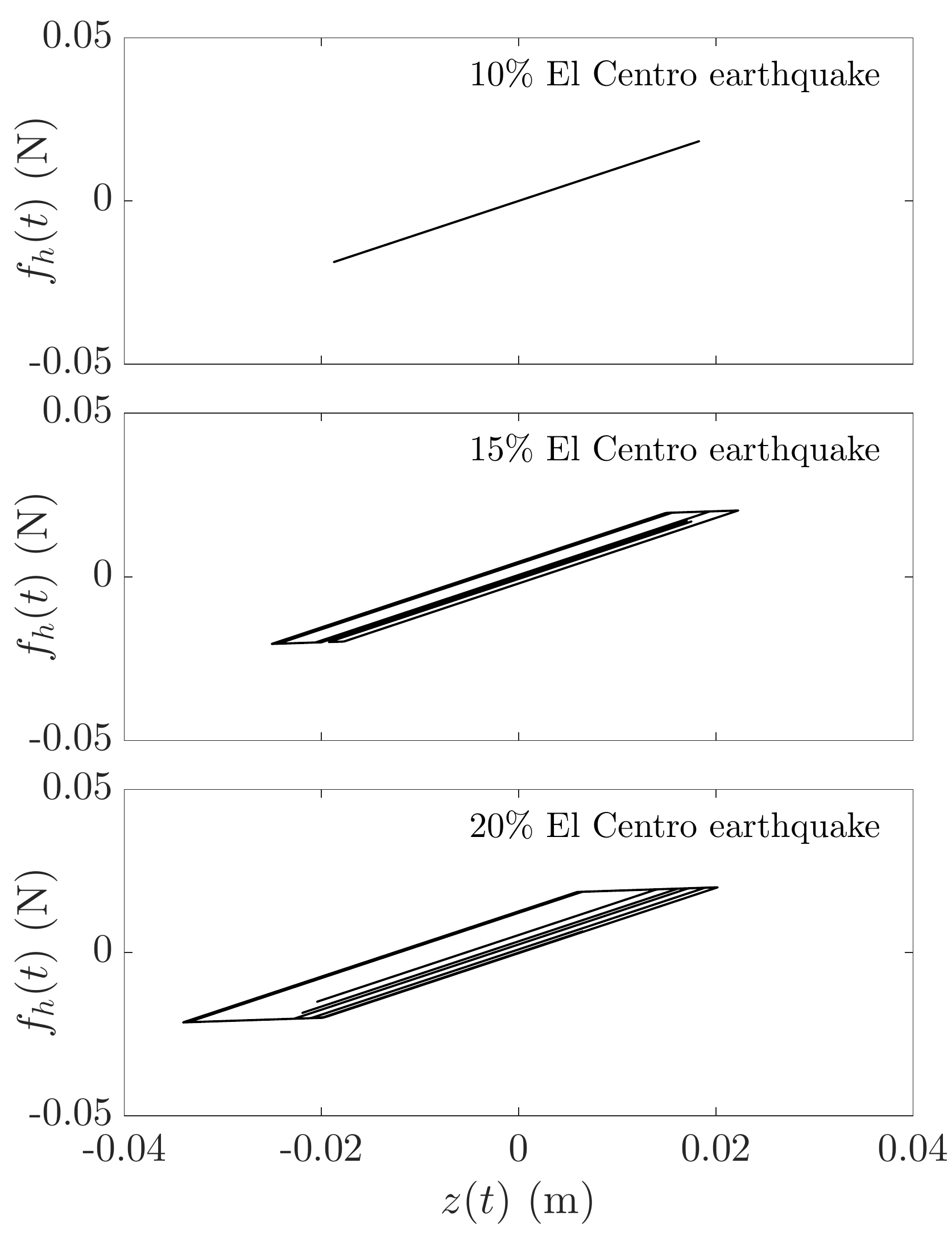}
		\label{fig:fig2b}
	\end{subfigure}
	\vspace*{-10pt}
	\caption{\emph{a) Oscillator response measurements for three levels of excitation; b) Oscillator hysteresis loops for the three levels of excitation (Example 1).}}
	\label{fig:fig2}
\end{figure}

\begin{table}
	\centering
	\ra{1}
	\caption{Posterior sample mean of parameter values for each model class representing the oscillator and for different levels of excitations (Example 1), UN denotes Unidentifiable parameter.}\label{tab:tab1}
	\begin{tabular*}{\columnwidth}{@{\extracolsep{\stretch{1}}}*{1}{l}*{6}{c}@{}}
		\toprule
		Excitation level& Model class& $c$(N.s/m)& $k_1$(N/m)& $k_2$(N/m)& $z_y$(cm)& $\sigma_v$(cm) \\ \midrule
		
		\multirow{3}{*}{$10\%$ El Centro}&$\mathcal{M}_1$&0.0199  &1.0008&   --- &   --- &0.0005\\
										 &$\mathcal{M}_2$&     ---&1.0036&   --- &  1.86 &0.0010\\
										 &$\mathcal{M}_3$&     ---&1.0034&   UN  &  1.84 &0.0010\\
		\multirow{3}{*}{$15\%$ El Centro}\rule{0pt}{3ex} &$\mathcal{M}_1$&0.0944 & 0.9661&  --- &   --- &0.0021\\
										 &$\mathcal{M}_2$&     ---&0.9978&   --- &  2.10 &0.0012\\
										 &$\mathcal{M}_3$&     ---&1.0141&0.1485 &  1.98 &0.0009\\
		\multirow{3}{*}{$20\%$ El Centro}\rule{0pt}{3ex} &$\mathcal{M}_1$&0.1978 & 0.9029&  --- &   --- &0.0061\\
										 &$\mathcal{M}_2$&     ---&0.9517&    ---&  2.26 &0.0039\\
										 &$\mathcal{M}_3$&     ---&1.0141&0.0950 &  1.97 &0.0008\\
		\bottomrule
	\end{tabular*}
\end{table}

Three sets of data are studied here with $10$, $15$, and $20\%$ scaling of the $1940$ El Centro earthquake record (north-south component) as the excitation corresponding to each data set. The earthquake excitation and displacement of the oscillator are measured for $40$ s with sampling frequency of $60$ Hz to give $N=2400$ data points. These synthetic data are contaminated with zero-mean Gaussian discrete white noise in which the variance $\sigma_v^2$ for each data set is selected so that the output prediction error gives a $5\%$ RMS noise-to-signal ratio over the associated displacement data from the actual system. This gives $\sigma_v$ equal to $4.6\times10^{-4}$ cm, $5.9\times10^{-4}$ cm, and $7.1\times10^{-4}$ cm for data sets generated using $10$, $15$, and $20\%$ scaling of the $1940$ El Centro earthquake record, respectively. The left panel of Figure \ref{fig:fig2} demonstrates the synthetic measurements for the three levels of excitation and the right panel of Figure \ref{fig:fig2} indicates the corresponding hysteresis loops. This figure shows that the dynamics of the structure is linear for $10\%$ scaling, mildly nonlinear for $15\%$ scaling, and strongly nonlinear for $20\%$ scaling of the El Centro earthquake record.

Here, we consider three different model classes: Model class $\mathcal{M}_1$ is a linear oscillator in which the uncertain model parameter vector $\boldsymbol{\theta}_s=[k_1,\,c]$, consisting of the elastic stiffness $k_1$ and viscous damping $c$. Model class $\mathcal{M}_2$ is a bilinear hysteretic oscillator with post-yield stiffness $k_2=0$ (elastoplastic oscillator) and no viscous damping $c=0$. Thus, the model parameter vector $\boldsymbol{\theta}_s=[k_1,\,z_y]$, consisting of the elastic stiffness $k_1$ and yield displacement $z_y$. Model class $\mathcal{M}_3$ is a bilinear hysteretic oscillator with no viscous damping $c=0$. In this model class, the uncertain model parameter vector $\boldsymbol{\theta}_s=[k_1,\,k_2,\,z_y]$, consisting of the elastic stiffness $k_1$, post-yield stiffness $k_2$, and yield displacement $z_y$. Note that none of these model classes match the exact model class used to generate the data. The prior distributions for the model parameters $k_1$, $k_2$, $c$, $z_y$ are chosen as independent uniform distributions over the intervals $(0,2)$ N/m, $(0,0.5)$ N/m, $(0,0.5)$ N.s/m, $(0,0.1)$ m, respectively.

\begin{figure}[htpb!]
	\centering
	\includegraphics[width=1\textwidth]{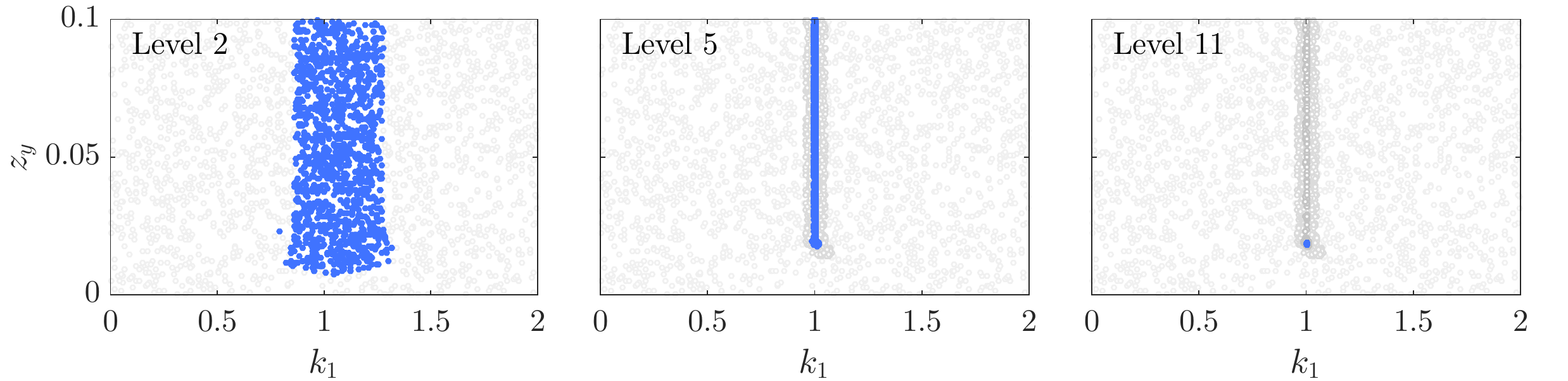}
	\caption{\emph{Scatter plot of 2000 samples in the $\{k_1,z_y\}$ space that are generated at levels 2, 5 and 11 (in blue) and their previous intermediate levels (in gray) of ABC-SubSim when updating model class $\mathcal{M}_2$ with data from the $10\%$ El Centro earthquake (Example 1).}}
	\label{fig:fig3}
\end{figure}

\begin{figure}[htpb!]
	\centering
	\includegraphics[width=1\textwidth]{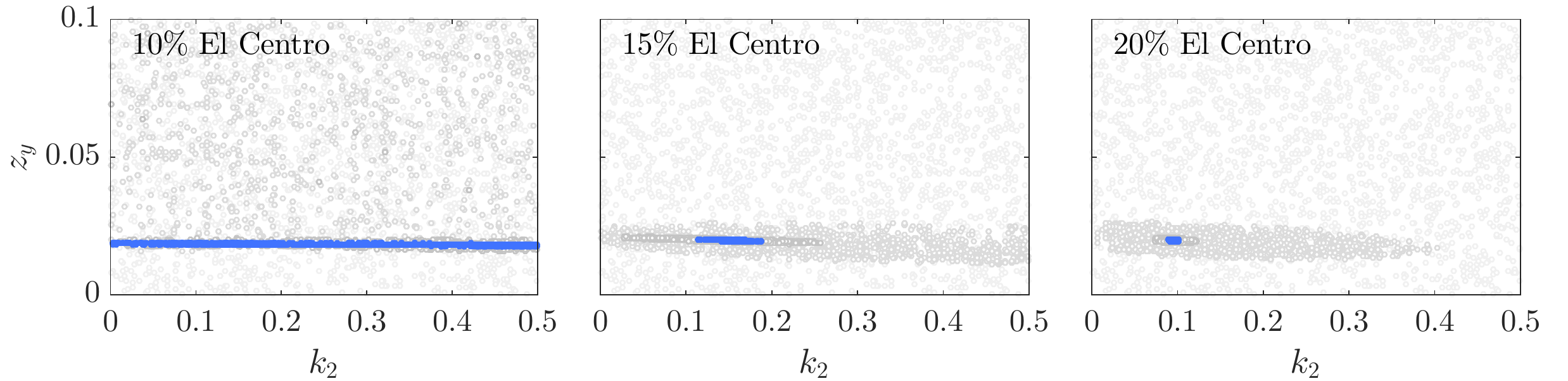}
	\caption{\emph{Scatter plot of 2000 posterior samples in the $\{k_2,z_y\}$ space when updating model class $\mathcal{M}_3$ for the intermediate levels (in gray) and the final level (in blue) and for different excitation levels (Example 1).}}
	\label{fig:fig4}
\end{figure}

The mean estimates of the model parameters for the three model classes and three different excitation levels are reported in Table \ref{tab:tab1}. The parameter estimates are the mean of 2000 posterior samples drawn by self-regulating ABC-SubSim. For model class $\mathcal{M}_1$ and for higher levels of excitation, Table \ref{tab:tab1} shows that the mean estimates have lower values of the linear stiffness and higher values of the damping coefficient to represent the hysteretic dissipated energy. Figure \ref{fig:fig3} shows the 2000 samples drawn in different levels of ABC-SubSim in the $\{k_1,z_y\}$ space when updating model class $\mathcal{M}_2$ using data from the $10\%$ El Centro earthquake record. As shown, the posterior samples are tightly clustered around the value $z_y=1.86$ cm. This might seem a counter-intuitive result since when the oscillator is excited with the $10\%$ El Centro earthquake, it behaves perfectly linear and so one expects $z_y$ to be unidentifiable. However, it seems that this value for $z_y$, which is slightly less than the maximum amplitude of the oscillations ($1.875$ cm), is an attempt to yield enough hysteretic dissipative energy to compensate for the lack of viscous damping in the model class $\mathcal{M}_2$. A similar behavior can be observed in Figure \ref{fig:fig4} for model class $\mathcal{M}_3$ where $z_y$ is pinned down around $1.84$ cm while $k_2$ is unidentifiable (because there is little post-yield response), which seems to be again an attempt to generate enough dissipative energy due to yielding to compensate for the lack of viscous damping. It can be observed in Figure \ref{fig:fig4} that for the higher levels of excitation, the structure experiences stronger nonlinear behavior and so the post-yielding stiffness parameter $k_2$ is pinned down more tightly. 


Table \ref{tab:tab2} presents the number of simulation levels $m$ and the final tolerance levels $\epsilon_{\scriptscriptstyle\mathcal{M}_j}$ which are adaptively selected by the algorithm to explore the posterior distribution of the parameters of different models $\mathcal{M}_j,\,j=1,\,2,\,3$, for the three levels of excitation. This table also shows the posterior probability of models $P(\mathcal{M}_j(\epsilon_{\scriptscriptstyle\mathcal{M}_j})|\mathcal{D}_N,\boldsymbol{M})$ obtained from (\ref{eq:eq26}) by evaluation of evidence (\ref{eq:eq27}) at the final tolerance levels $\epsilon_{\scriptscriptstyle\mathcal{M}_j}$ and equal prior probabilities $P(\mathcal{M}_j|\boldsymbol{M})=1/3$. According to this result, in the case of $10\%$ scaling of the $1940$ El Centro earthquake record where the structure behaves linearly (see Figure \ref{fig:fig2b}), $\mathcal{M}_1$ gives the largest posterior probability. For the higher levels of excitations, when the dissipated energy is dominated by the hysteretic dissipative energy, $\mathcal{M}_3$ is the most plausible model. 
This example reflects the important point made by the famous statement that ``all models are wrong, but some are useful" \cite{box1987empirical} and, furthermore, the ``best model" for the system analysis and response predictions depends on the data that is used for system identification.  

\begin{table}
	\centering
	\ra{1}
	\caption{Posterior probability of different model classes together with final tolerance level and number of simulation levels for three-story Masing building (Example 2).}\label{tab:tab2}
	\begin{tabular*}{\columnwidth}{@{\extracolsep{\stretch{1.5}}}*{1}{l}*{11}{c}@{}}
		\toprule
		& \multicolumn{3}{c}{$10\%$ El Centro}& &\multicolumn{3}{c}{$15\%$ El Centro}& &\multicolumn{3}{c}{$20\%$ El Centro}\\ \cmidrule{2-4} \cmidrule{6-8} \cmidrule{10-12}
		Model class & $\mathcal{M}_1$ & $\mathcal{M}_2$& $\mathcal{M}_3$& & $\mathcal{M}_1$ & $\mathcal{M}_2$& $\mathcal{M}_3$& &$\mathcal{M}_1$ & $\mathcal{M}_2$& $\mathcal{M}_3$ \\ \midrule
		
		Sim. levels ($m$)  & 9 & 9 & 9 & &7 & 9 & 10 & &5 & 7 & 12\\
		Tol. level ($\epsilon_{\scriptscriptstyle\mathcal{M}_j}$) &0.0007&0.0014&0.0014& &0.0029&0.0016&0.0012& &0.0084&0.0054&0.0011\\
		$P(\mathcal{M}_j(\epsilon_{\scriptscriptstyle\mathcal{M}_j})|\mathcal{D}_N,\boldsymbol{M})$  & 1 & 0 & 0 & & 0 & 0 & 1 & & 0 & 0 & 1\\
		\bottomrule
	\end{tabular*}
\end{table}

\begin{figure}[htpb!]
	\centering
	\includegraphics[width=1\textwidth]{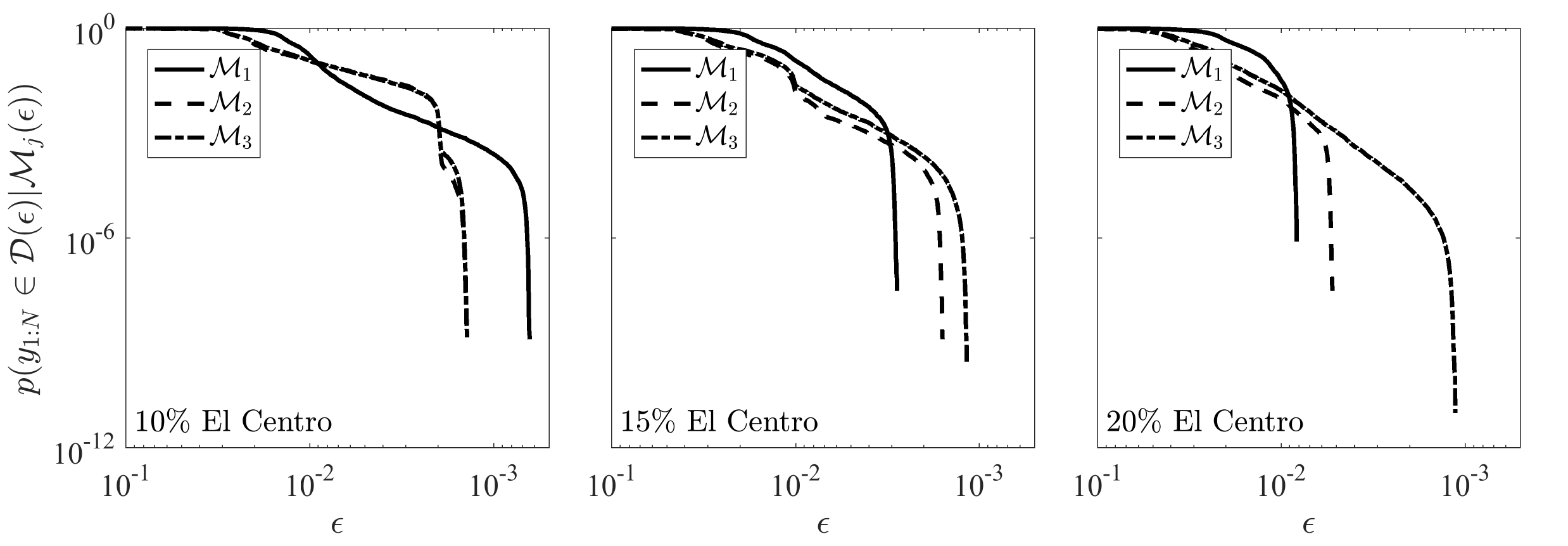}
	\caption{\emph{The probability of entering the data-approximating region $\mathcal{D}(\epsilon)$ against tolerance level $\epsilon$ (Example 1).}}
	\label{fig:fig5}
\end{figure}

\begin{figure}[htpb!]
	\centering
	\includegraphics[width=1\textwidth]{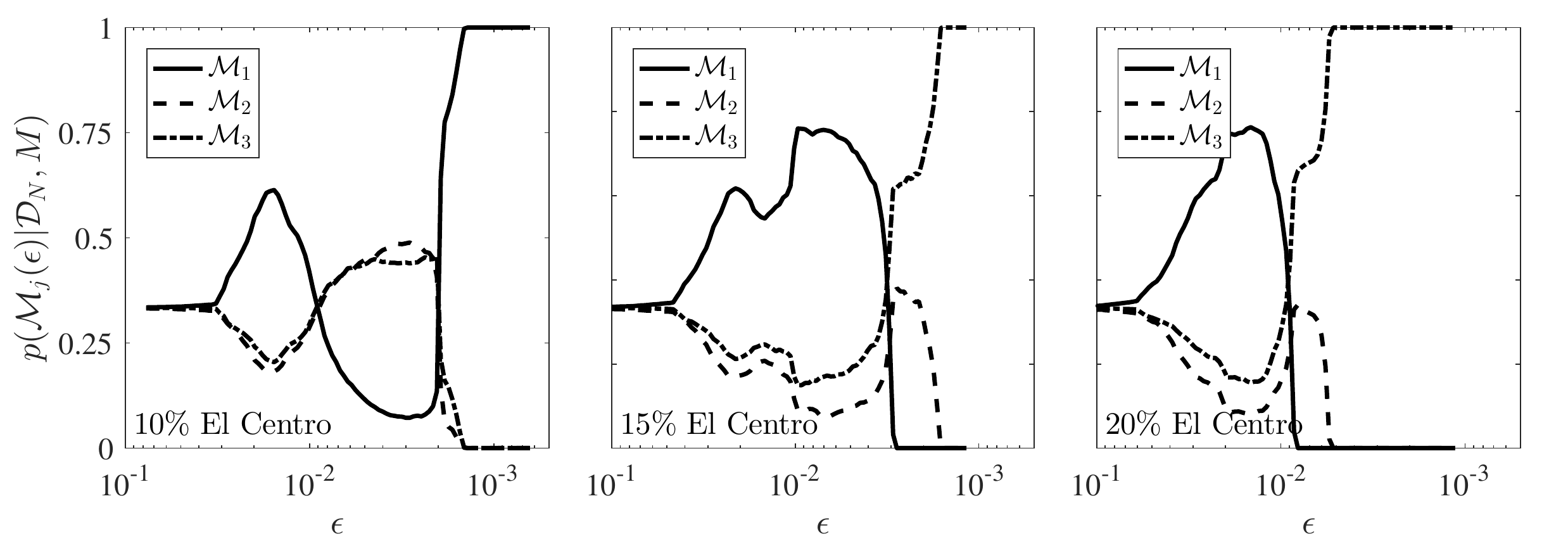}
	\caption{\emph{The posterior probability of different model classes $\mathcal{M}_j$ against tolerance level $\epsilon$ (Example 1).}}
	\label{fig:fig6}
\end{figure}

One of the difficulties for any ABC algorithm is to select the final tolerance level for which the self-regulating ABC-SubSim algorithm brings a straightforward solution, as explained in Remark \ref{rmk:4}. The agreement between the approximate posterior probabilities $P(\mathcal{M}_j(\epsilon_{\scriptscriptstyle\mathcal{M}_j})|\mathcal{D}_N,\boldsymbol{M})$ presented in Table \ref{tab:tab2} with those reported by Beck and Yuen \cite{beck2004model} shows the validity of the stopping criterion used in self-regulating ABC-SubSim. To further emphasize the importance of choosing an appropriate final tolerance level, the probability that $\boldsymbol{y}_{1:N}$ lies in the ball of radius $\epsilon$ around the data vector $\hat{\boldsymbol{z}}_{1:N}$ and the posterior probability $P(\mathcal{M}_j(\epsilon)|\mathcal{D}_N,\boldsymbol{M})$ for different model classes are depicted versus the tolerance level $\epsilon$ in Figures \ref{fig:fig5} and \ref{fig:fig6}, respectively. As $\epsilon$ goes down from $0.1$ to $\epsilon_{\scriptscriptstyle\mathcal{M}_j}$, $P(\mathcal{M}_j(\epsilon)|\mathcal{D}_N,\boldsymbol{M})$ varies between the model prior probabilities at $\epsilon=0.1$ and the true model posterior probabilities at $\epsilon_{\scriptscriptstyle\mathcal{M}_j}$.

\subsection*{Example 2: Three-story Masing shear-building under seismic excitation}
\begin{figure}[htpb!]
	\centering
	\includegraphics[width=.32\textwidth]{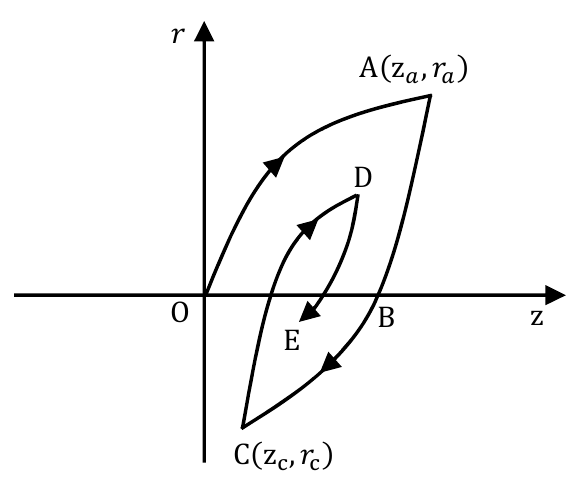}
	\caption{\emph{Hysteresis loops for transient loading of the extended Masing model \cite{jayakumar1988system}.}}
	\label{fig:fig7}
\end{figure}

The second example is taken from Muto and Beck \cite{muto2008bayesian} where the Transitional Markov Chain Monte Carlo (TMCMC) algorithm has been used for Bayesian updating and model selection of the class of Masing hysteretic structural models. This example considers a three-story shear building with the following equation of motion:
\begin{equation}\label{eq:eq44}
	\boldsymbol{M}\ddot{\boldsymbol{z}}(t)+\boldsymbol{C}\dot{\boldsymbol{z}}(t)+\boldsymbol{f}_h=-\boldsymbol{M1}u(t)
\end{equation}
	where $\boldsymbol{z}(t)\in\mathbb{R}^3$ is the horizontal displacement vector relative to the ground; $\boldsymbol{M}$, $\boldsymbol{C}\in \mathbb{R}^{3\times3}$ are the mass and damping matrices; $u(t)$ is the horizontal ground acceleration; and $\boldsymbol{1}=[1\,\,1\,\,1]^T.$ The restoring force for the $i$th story is given by:
\begin{equation}\label{eq:eq45}
	f_{h,i}=r_i-r_{i+1}
\end{equation}
where the inter-story shear force-deflection relation is given by the differential equation: 
\begin{equation}\label{eq:eq46}
	\dot{r}_i=k_i(\dot{z}_i-\dot{z}_{i-1})\bigg[1-\bigg|\frac{r_i}{r_{u,i}}\bigg|^{\alpha_i}\bigg]
\end{equation}
Here, $k_i$ is the small-amplitude inter-story stiffness, $r_{u,i}$ is the story ultimate strength and $\alpha_i$ is the elastic-to-plastic transition parameter. The force-deflection relationship given in (\ref{eq:eq46}) defines the initial loading curve OA in Figure \ref{fig:fig7}. Any other loading curve can be selected according to two \textit{extended Masing rules} \cite{jayakumar1988system,jayakumar1987modeling,ashrafi2007generalized}:
\begin{enumerate}
	\item The force-deflection relation for any loading curve other than the initial loading (\ref{eq:eq46}) is described by the differential equation:
	\begin{equation}\label{eq:eq47}
	\dot{r}_i=k_i(\dot{z}_i-\dot{z}_{i-1})\bigg[1-\bigg|\frac{r_i-r^*}{2r_{u,i}}\bigg|^{\alpha_i}\bigg]
	\end{equation}
	where $r^*$ is the restoring force at the latest load reversal point. For instance, (\ref{eq:eq47}) gives the loading curve AC in Figure \ref{fig:fig7} if $r^*=r_a$.
	\item Once an interior loading curve crosses a curve from a previous load cycle, the load deformation continues that of the previous cycle. For instance, if the curve DE is continued to point C, it follows the force-deflection relation of curve ABC. 
\end{enumerate}

It should be noted that a wide variety of hysteretic models can be described by using the two extended Masing rules through the choice of the initial load curve. Thus, the class of Masing hysteretic model with restoring force-deflection relation (\ref{eq:eq46}) used here is only a special class of Masing models. 

In this example, the structure has a known story mass of $1.25\times10^5$ kg and a known earthquake ground acceleration defining the excitation $u(t)$. Similar to \cite{muto2008bayesian}, the east-west component of the Sylmar ground-motion record from the County Hospital Parking Lot during 1994 Northridge earthquake in California is used here as the excitation. The actual values for the model parameters for each story $i=1,2,3$ are: small-amplitude inter-story stiffnesses $k_i=2.5\times10^8$ N/m, ultimate strengths $r_{u,i}=1.75\times10^6$ N, and elastic-to-plastic transition parameters $\alpha_i=4$. The viscous damping matrix $\boldsymbol{C}$ is modeled using Rayleigh damping $\boldsymbol{C}=c_M\boldsymbol{M}+c_K\boldsymbol{K}$ with coefficients $c_M=0.293$ and $c_K=2.64\times10^{-4}$. This set of structural parameters gives the three small-amplitude natural frequencies as $\omega_1=3.17$ Hz, $\omega_2=8.88$ Hz and $\omega_3=12.83$ Hz, and the modal damping ratios as $\zeta_1= \zeta_2=0.01$ and $\zeta_3=0.012$.

In this study, we use the fixed-point exponential integrator \cite{chen2016efficient} together with the particular numerical implementation of the two extended Masing rules that was proposed in \cite{thyagarajan1989modeling} to simulate samples of the response time history $\boldsymbol{y}_{1:N}$ for given values of the uncertain parameters and the Sylmar ground-motion record. 

\begin{figure}[htpb!]
	\centering
	\includegraphics[width=1\textwidth]{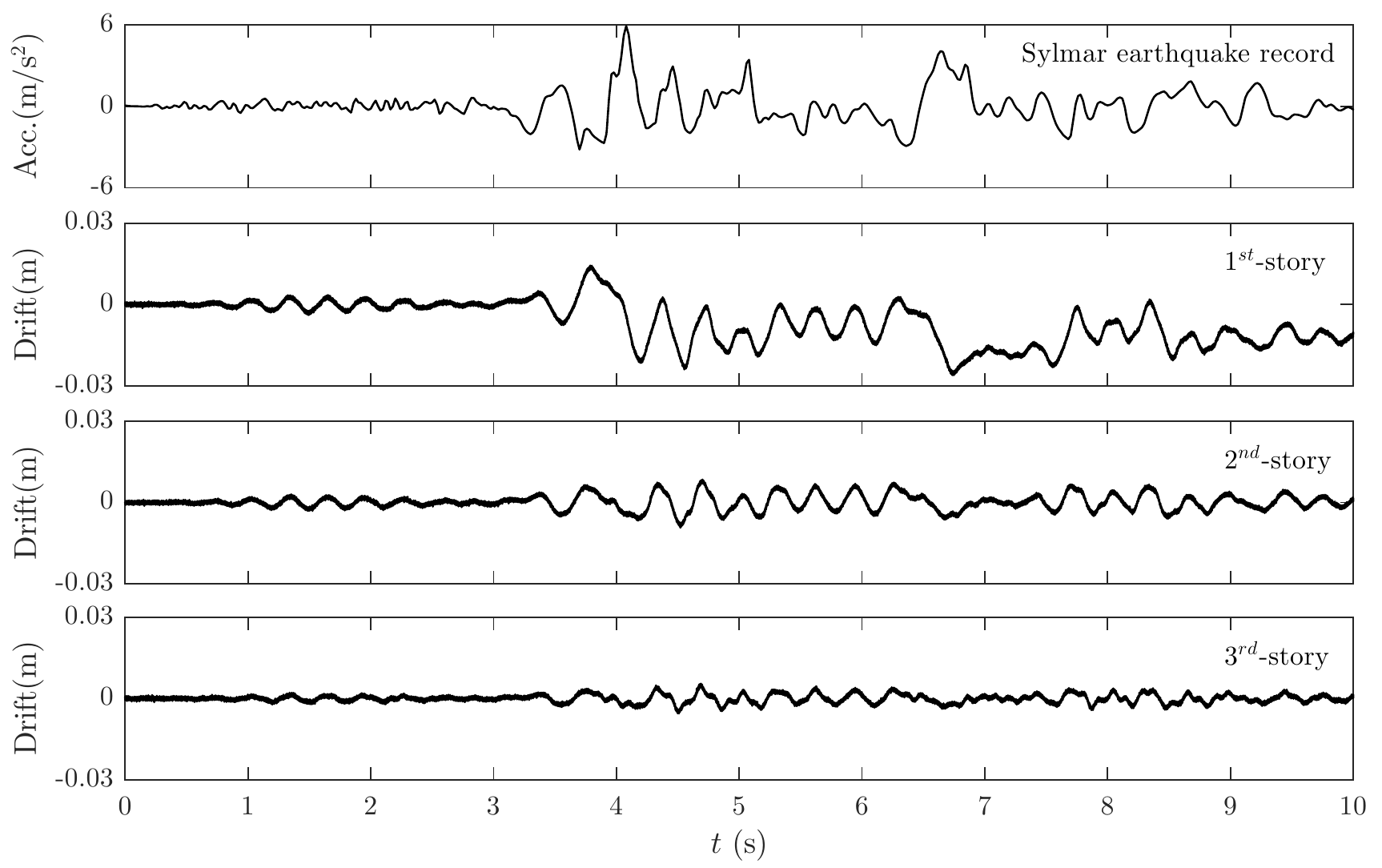}
	\caption{\emph{Inter-story drift time histories and the Sylmar ground-motion record (Example 2).}}
	\label{fig:fig8}
\end{figure}

\begin{figure}[htpb!]
	\centering
	\includegraphics[width=1\textwidth]{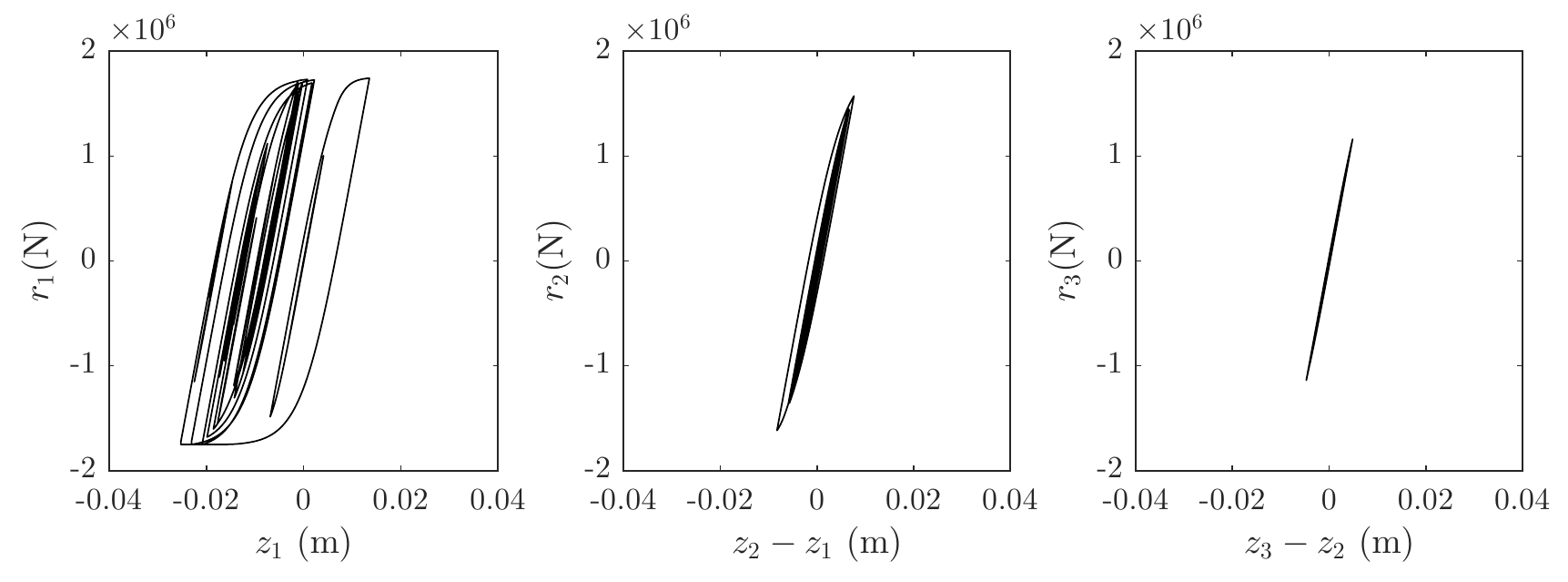}
	\caption{\emph{Simulated inter-story restoring forces against inter-story drifts (Example 2).}}
	\label{fig:fig9}
\end{figure}

The synthetic response data for system identification is the inter-story drift time histories when the uncertain parameters are set to their actual values and the standard deviation of the uncertain output error is set to $0.03$ cm to give a $5\%$ RMS noise-to-signal ratio. The simulated data, which are shown in Figure \ref{fig:fig8}, correspond to $500$ data points from each story with a sampling rate of 0.02 s. Figure \ref{fig:fig9} shows the hysteresis loops for each story generated by the Sylmar ground motion. This figure indicates that the simulated structure experiences strong yielding in the first story and moderate yielding in the second story whereas it exhibits almost linear behavior in the top story.


Four model classes are studied for system identification. For all model classes, the story masses are taken as known and set to their actual values. Generally, the two parameters, $c_M$ and $c_K$, specifying the viscous damping matrix and the nine parameters of the hysteresis model $k_i$, $r_{u,i}$ and $\alpha_i$, $i=1,2,3$ in (\ref{eq:eq46}), make the vector of uncertain structural model parameters $\theta_s$. For model classes $\mathcal{M}_1$ and $\mathcal{M}_2$, the elastic-to-plastic transition parameters are constrained to be equal for all three stories whereas they are allowed to vary for model classes $\mathcal{M}_3$ and $\mathcal{M}_4$. The model classes $\mathcal{M}_1$ and $\mathcal{M}_3$ contain no viscous damping, but model classes $\mathcal{M}_2$ and $\mathcal{M}_4$ do and the Rayleigh damping coefficients $c_M$ and $c_K$ are estimated for these model classes. Therefore, in contrast to Example 1, a subset of the candidate model classes, i.e., $\mathcal{M}_2$ and $\mathcal{M}_4$, contains the model used to generate the data. 

The prior distribution over the nine-dimensional parameter space of the hysteresis model is selected to be the product of nine lognormal PDFs with logarithmic mean value of $\log(2.5\times10^8)$ for $k_i,\,i=1,2,3$, $\log(2.5\times10^6)$ for $r_{u,i},\,i=1,2,3$ and $\log(4)$ for $\alpha_i,\,i=1,2,3$ and a logarithmic standard deviation of $0.5$ for all of them. The prior distributions for the parameters of the viscous damping matrix are defined as independent uniform PDFs over the interval $[0,1.5]$ for $c_M$ and $[0,1.5\times10^{-3}]$ for $c_K$.

\begin{figure}[htpb!]
	\centering
	\includegraphics[width=1\textwidth]{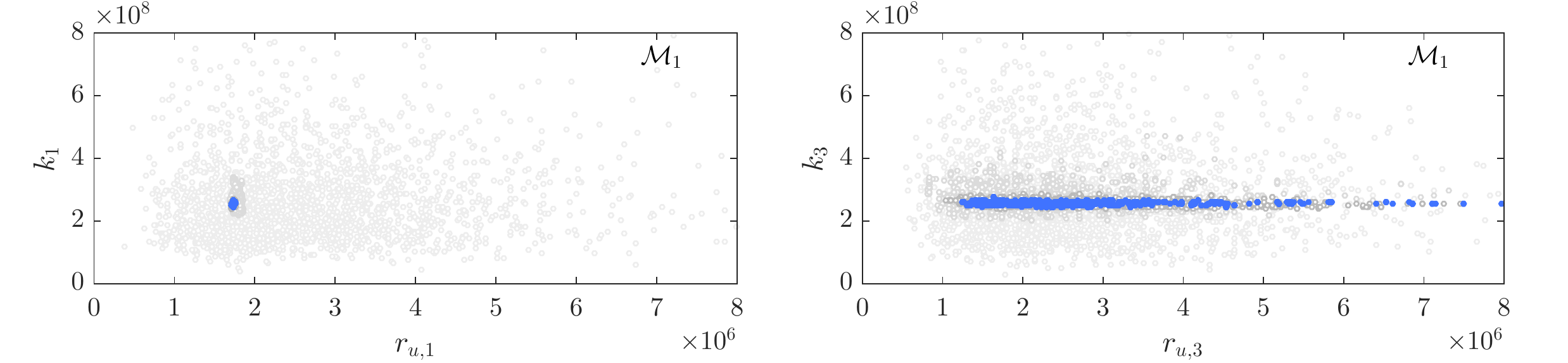}
	\caption{\emph{Scatter plot of 2000 posterior samples plotted in $\{r_{u,1},\alpha_1\}$ (left) and $\{r_{u,3},\alpha_3\}$ (right) spaces when updating model class $\mathcal{M}_1$ for some intermediate levels (in gray) and the final level (in blue) (Example 2).}}
	\label{fig:fig10}
\end{figure}

\begin{figure}[htpb!]
	\centering
	\includegraphics[width=1\textwidth]{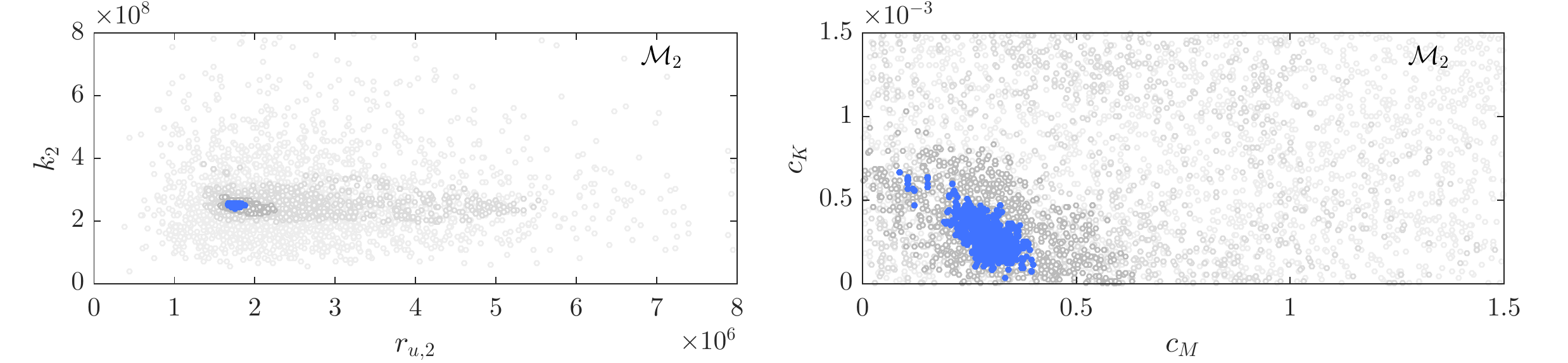}
	\caption{\emph{Scatter plot of 2000 posterior samples plotted in $\{r_{u,2},\alpha_2\}$ (left) and $\{c_M,c_K\}$ (right) spaces when updating model class $\mathcal{M}_2$ for some intermediate levels (in gray) and the final level (in blue) (Example 2).}}
	\label{fig:fig11}
\end{figure}

\begin{figure}[htpb!]
\centering
\includegraphics[width=1\textwidth]{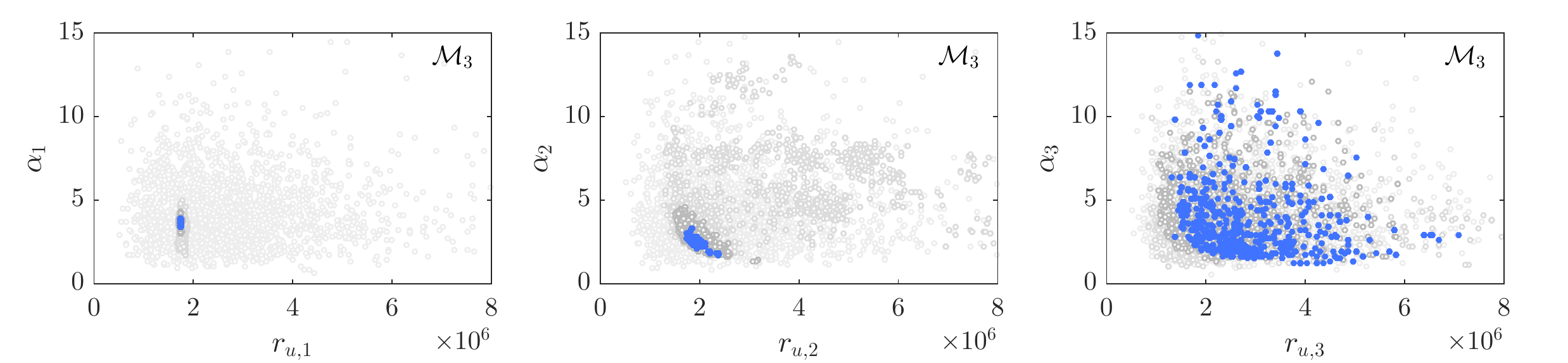}
\caption{\emph{Scatter plot of 2000 posterior samples plotted in $\{r_{u,1},\alpha_1\}$ (left) and $\{r_{u,2},\alpha_2\}$ (middle) and $\{r_{u,3},\alpha_3\}$ (right) spaces when updating model class $\mathcal{M}_3$ for some intermediate levels (in gray) and the final level (in blue) (Example 2).}}
\label{fig:fig12}
\end{figure}

\begin{figure}[htpb!]
\centering
\includegraphics[width=1\textwidth]{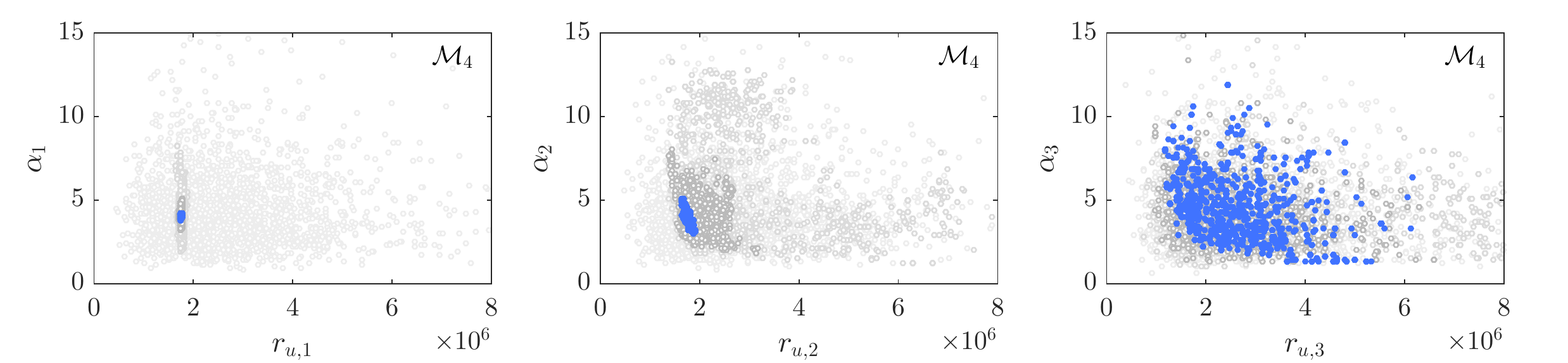}
\caption{\emph{Scatter plot of 2000 posterior samples plotted in $\{r_{u,1},\alpha_1\}$ (left) and $\{r_{u,2},\alpha_2\}$ (middle) and $\{r_{u,3},\alpha_3\}$ (right) spaces when updating model class $\mathcal{M}_4$ for some intermediate levels (in gray) and the final level (in blue) (Example 2).}}
\label{fig:fig13}
\end{figure}

Table \ref{tab:tab3} shows the MAP (maximum a posteriori) values and the standard deviations of the uncertain parameters obtained for all model classes. The MAP value for each parameter is estimated by fitting a lognormal PDF to the samples drawn from their posterior distribution meaning that this is different from the posterior sample mean. Figures \ref{fig:fig10}-\ref{fig:fig13} show $2000$ samples obtained from the self-regulating ABC-SubSim algorithm for some of the uncertain parameters of the four model classes.

For model classes $\mathcal{M}_1$ and $\mathcal{M}_2$, the posterior samples of the small-amplitude stiffnesses are tightly clustered (see Figures \ref{fig:fig10} and \ref{fig:fig11}). The posterior samples of the ultimate strength for the first story is well-constrained, for the second story they show slightly higher level of uncertainty and for the third story they exhibit a high level of uncertainty (see Figures \ref{fig:fig10} and \ref{fig:fig11}). This phenomenon can be understood by looking at Figure \ref{fig:fig9}, which demonstrates that a noticeable yielding occurred in the first story. This means that there is enough information in the response data to estimate the ultimate strength for the first story. On the other hand, this figure shows a nearly linear behavior for the third story, so the response data only impose a lower bound on the ultimate strength. These results are very similar to those reported by Muto and Beck \cite{muto2008bayesian}. For model class $\mathcal{M}_2$, the posterior distribution for the parameters of the viscous damping matrix reveals a high level of uncertainty. Presumably, this can be attributed to the fact that the response is less sensitive to the variation of the parameters of the hysteresis model.

\begin{table}
	\centering
	\ra{1}
	\caption{The maximum a posteriori parameter values and the standard deviations (in parentheses) obtained from fitting a lognormal distribution to the posterior samples (Example 2).}\label{tab:tab3}
	\begin{tabular*}{\columnwidth}{@{\extracolsep{\stretch{1}}}*{1}{l}*{4}{c}@{}}
		\toprule
		Model class & $\mathcal{M}_1$ & $\mathcal{M}_2$& $\mathcal{M}_3$& $\mathcal{M}_4$\\ \midrule
		
		$k_1\,(10^8\, N/m)$  &2.586 (0.026)&2.497 (0.018)&	2.545 (0.029)&	2.490 (0.015)\\
		$k_2\,(10^8\, N/m)$  &2.455 (0.044)&2.499 (0.025)&	2.539 (0.042)&	2.509 (0.034)\\
		$k_3\,(10^8\, N/m)$  &2.566 (0.054)&2.490 (0.023)&	2.545 (0.061)&	2.504 (0.025)\\
		$r_{u,1}\,(10^6\, N)$&1.737 (0.004)&1.749 (0.003)&	1.746 (0.006)&	1.751 (0.003)\\
		$r_{u,2}\,(10^6\, N)$&1.779 (0.064)&1.750 (0.037)&  1.924 (0.152)&	1.757 (0.054)\\
		$r_{u,3}\,(10^6\, N)$&2.056 (1.014)&2.140 (0.772)&	2.358 (0.934)&	2.154 (1.083)\\
		$\alpha_1$			 &3.430 (0.090)&3.981 (0.075)&	3.447 (0.145)&	4.041 (0.094)\\
		$\alpha_2$			 &$=\alpha_1$  &$=\alpha_1$	 &  2.626 (0.300)&	3.863 (0.411)\\
		$\alpha_3$			 &$=\alpha_1$  &$=\alpha_1$	 &  2.552 (2.607)&	3.332 (2.047)\\
		$c_M (s^{-1})$		 &	---		   &0.259 (0.071)&	---		     &	0.283 (0.069)\\
		$c_K (10^{-4}s)$	 &  ---		   &2.295 (1.322)&	---	         &  2.116 (0.909)\\
		$\sigma_v (10^{-4}m)$&5.293 (0.063)&3.197 (0.064)&	5.163 (0.084)&	3.176 (0.048)\\
		
		\bottomrule
	\end{tabular*}
\end{table}

For model classes $\mathcal{M}_3$ and $\mathcal{M}_4$, the posterior distribution for the small-amplitude stiffnesses are also compactly clustered but they are not graphically shown here (see Table \ref{tab:tab3}). Both model classes exhibit an almost similar behavior in the parameter space $\{r_{u,i},\alpha_i\}$ (see Figures \ref{fig:fig12} and \ref{fig:fig13}). As expected, the parameters for the first story, $r_{u,1}$ and $\alpha_1$, are globally identifiable for both models. For the third story, the posterior distribution of the parameters $r_{u,3}$ and $\alpha_3$ shows a large spread over the parameter space with a clear lower bound. This can be attributed to the fact that the third story does not experience yielding and the lower bound is the only information that can be extracted from the data. However, the joint posterior distribution of $r_{u,2}$ and $\alpha_2$ in model class $\mathcal{M}_3$ differs from its counterpart in model class $\mathcal{M}_4$ (see Figures \ref{fig:fig12} and \ref{fig:fig13}). The lack of viscous damping in model class $\mathcal{M}_3$ apparently forces the posterior samples in $\{\alpha_2,r_{u,2}\}$ space to be clustered in a region around a lower value of $\alpha_2$ and a higher value for $r_{u,2}$, so that the dissipated hysteretic energy can compensate for the lack of viscous damping. In model class $\mathcal{M}_4$, the estimated Rayleigh damping parameters are rather close to their actual values and so the need for a higher hysteretic dissipation energy is mitigated, explaining why the posterior samples in $\{\alpha_2,r_{u,2}\}$ are clustered around their actual values. We note that the results presented for models $\mathcal{M}_3$ and $\mathcal{M}_4$ are to some extent different from their counterparts reported by Muto and Beck \cite{muto2008bayesian}. This difference can be explained by the fact that simulating the response time history $\boldsymbol{y}_{1:N}$ from a structure with hysteretic restoring forces is very dependent on the numerical schemes, e.g., type of time integrator used for the numerical implementations, and since the schemes used in this study are different from those of their study, results can be expected to be different to some extent.

Table \ref{tab:tab4} shows the number of simulation levels $m$ and the final tolerance levels $\epsilon_{\scriptscriptstyle\mathcal{M}_j}$ for different model classes. This table also presents the posterior probability of model classes $P(\mathcal{M}_j(\epsilon_{\scriptscriptstyle\mathcal{M}_j})|\mathcal{D}_N,\boldsymbol{M}),\,j=1,2,3,4$ calculated from (\ref{eq:eq26}) by evaluation of evidence (\ref{eq:eq27}) at the final tolerance levels $\epsilon_{\scriptscriptstyle\mathcal{M}_j}$ and equal prior probabilities $P(\mathcal{M}_j|\boldsymbol{M})=1/4$ for the models. It is not surprising that the posterior probability for the model classes favors model class $\mathcal{M}_2$ since it contains the model used to generate the synthetic data and has two parameters less than model class $\mathcal{M}_4$, which also contains the data-generating model. As shown by the information-theoretic expression for the log evidence in \cite{muto2008bayesian}, the posterior probability of a model class is controlled by a trade-off between the posterior average data fit (the posterior mean of the log-likelihood) and the amount of information extracted from data (the relative entropy of the posterior with respect to the prior). $\mathcal{M}_2$ and $\mathcal{M}_4$ give essentially the same average data fit but $\mathcal{M}_2$ extracts less information abouts its parameters. 


\begin{table}
	\centering
	\ra{1}
	\caption{Posterior probability of different model classes together with final tolerance level and number of simulation levels for three-story Masing building (Example 2).}\label{tab:tab4}
	\begin{tabular*}{\columnwidth}{@{\extracolsep{\stretch{1}}}*{1}{l}*{4}{c}@{}}
		\toprule
		Model class & $\mathcal{M}_1$ & $\mathcal{M}_2$& $\mathcal{M}_3$& $\mathcal{M}_4$\\ \midrule
		
		Sim. levels ($m$) 				   &       10          &   		10		&	11			&	12\\
		Tol. level ($\epsilon_{\scriptscriptstyle\mathcal{M}_j}$)  			   &6.80$\times10^{-4}$&4.25$\times10^{-4}$&7.10$\times10^{-4}$&4.25$\times10^{-4}$\\
		$P(\mathcal{M}_j(\epsilon_{\scriptscriptstyle\mathcal{M}_j})|\mathcal{D}_N,\boldsymbol{M})$  &        0          &     0.982		&	0			& 0.018\\
		\bottomrule
	\end{tabular*}
\end{table}

\begin{figure}[htbp!]
	\begin{subfigure}[b]{0.5\columnwidth}
		\caption{}
		\vspace*{-7pt}
		\centering
		\includegraphics[scale=0.5]{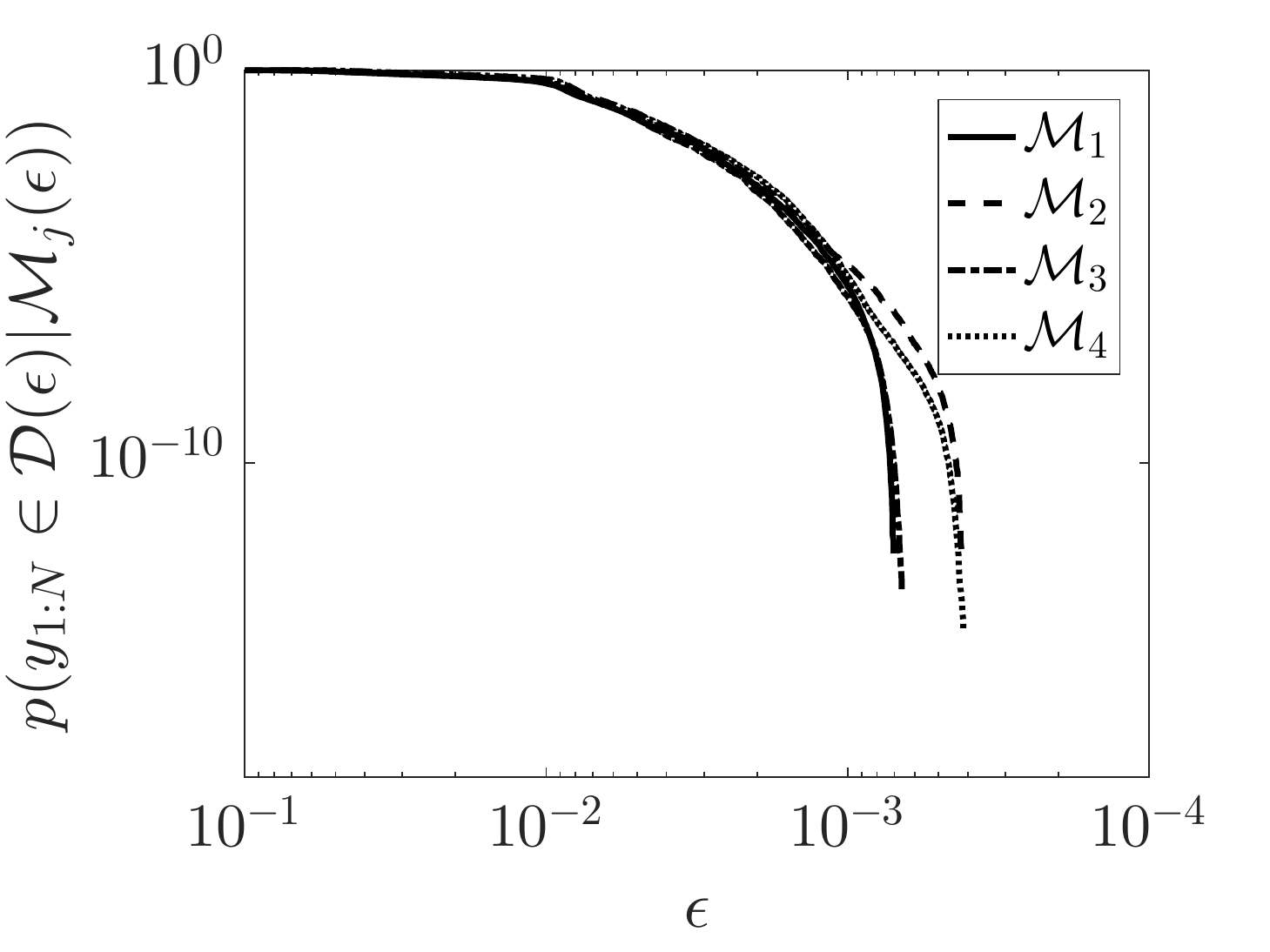}
		\label{fig:fig14a}
	\end{subfigure}
	\begin{subfigure}[b]{0.5\columnwidth}
		\caption{}
		\vspace*{-7pt}
		\centering
		\includegraphics[scale=0.5]{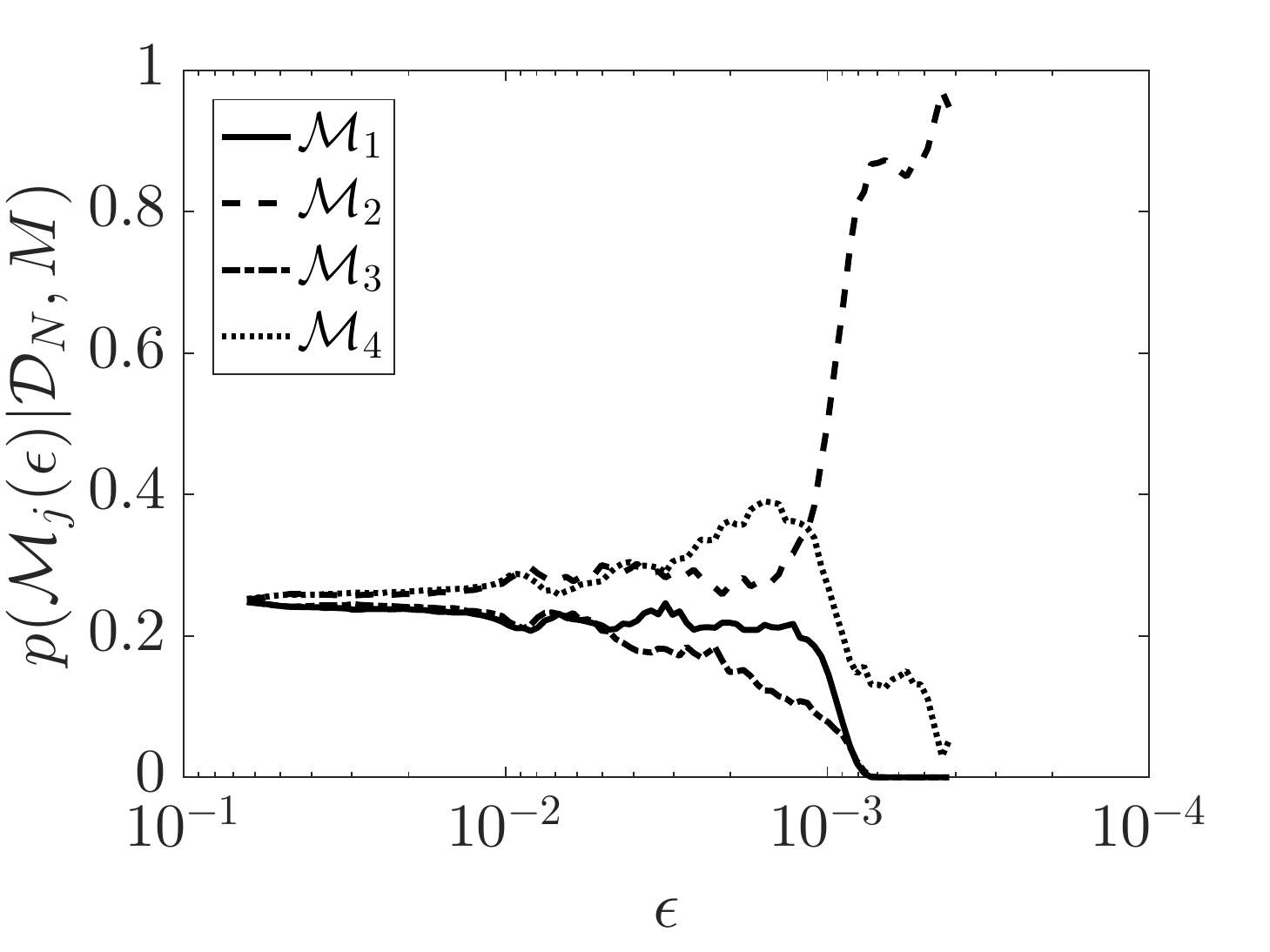}
		\label{fig:fig14b}
	\end{subfigure}
	\vspace*{-15pt}
	\caption{\emph{a) The probability of entering the data-approximating region $\mathcal{D}(\epsilon)$ against tolerance level $\epsilon$; b) The posterior probability of different model classes $\mathcal{M}_j$ against tolerance level $\epsilon$ (Example 2).}}
	\label{fig:fig14}
\end{figure}

Similar to the previous example, the approximate posterior probabilities $P(\mathcal{M}_j(\epsilon_{\scriptscriptstyle\mathcal{M}_j})|\mathcal{D}_N,\boldsymbol{M})$ presented in Table \ref{tab:tab4} are in agreement with those reported by Muto and Beck \cite{muto2008bayesian} which shows that the self-regulating ABC-SubSim algorithm selected proper values for the final tolerance levels $\epsilon_{\scriptscriptstyle\mathcal{M}_j}$. Figures \ref{fig:fig14}(a and b), respectively show the probability that $\boldsymbol{y}_{1:N}$ falls in the data-approximating region $\mathcal{D}(\epsilon)$ and the posterior probability $P(\mathcal{M}_j(\epsilon)|\mathcal{D}_N,\boldsymbol{M})$ for different model classes versus the tolerance level $\epsilon$. As $\epsilon$ goes down from $0.1$ to $\epsilon_{\scriptscriptstyle\mathcal{M}_j}$, $P(\mathcal{M}_j(\epsilon)|\mathcal{D}_N,\boldsymbol{M})$ varies between the model prior probabilities at $\epsilon=0.1$ and the true model posterior probabilities at $\epsilon_{\scriptscriptstyle\mathcal{M}_j}$. 

It is worth noting that both the parameter vector and the tolerance level are taken to be the same across models in the traditional ABC approach to model comparison explained in Algorithm \ref{algorithm:2} \cite{ratmann2009model}. This makes the estimates of the posterior probability of model classes sensitive to (\textit{i}) the proposal PDF of the Markov chain used within the sampling algorithm \cite{marin2012approximate}, and (\textit{ii}) the choice of a final tolerance level $\epsilon$. The former dependency should not occur since is not related to the inference problem under study. The choice of a unique tolerance level $\epsilon$ that works across all models is very delicate, since, as illustrated in Figure \ref{fig:fig6}, a wrong choice of $\epsilon$ can result in a significant bias in the ABC approximation of the model posterior probabilities. However, the proposed model selection procedure which is based on the hierarchical state-space formulation of dynamic models and the self-regulating ABC-SubSim algorithm alleviates these type of difficulties by independently estimating the model evidence for each of the models under comparison. 

\section{Concluding remarks}
In the current state of the art, ABC methods can only be used for model class selection in a very limited range of models for which a set of sufficient summary statistics can be found so that it also guarantees sufficiency across the set of models under study. In this paper, a new ABC model selection procedure has been presented which broadens the realm of ABC-based model comparison to be able to assess dynamic models. In the proposed procedure, a dynamic problem is formulated in terms of a general hierarchical state-space model such that the normalizing constant associated to its exact posterior distribution using the entire data provides an unbiased estimator of the model evidence as an error tolerance level $\epsilon\rightarrow0$. 

The self-regulating ABC-SubSim provides a straightforward way to estimate the model evidence and, as a result, the posterior probability of models as a function of the error tolerance level $\epsilon$. This enables us to better understand the model choices made in the earlier applications of the ABC-based model comparison methods. Furthermore, a new solution based on the Laplace's method of asymptotic approximations is presented to mitigate the fundamental difficulty of the ABC algorithms to learn the parameters specifying the uncertain state and output prediction errors in a stochastic state-space model. It has the key advantage that the approximated marginal distribution of the model parameters is insensitive to the prior adopted for the uncertain prediction error variance. 

Two illustrative examples with synthetic data are selected from the Bayesian system identification literature to show the estimation of the model class evidences and posterior probabilities obtained by the self-regulating ABC-SubSim algorithm. The first example shows the successful application of the self-regulating ABC-SubSim for Bayesian model class selection when the true system is not among the competing model classes. The second example shows the capability of the self-regulating ABC-SubSim algorithm to efficiently explore a posterior distribution with a relatively high-dimensional parameter space.    

\section*{Acknowledgment}
The first author of this paper wants to express his gratitude to the California Institute of Technology (Caltech) for kindly hosting him during the course of this work.

\section*{References}

\bibliography{ABC}

\end{document}